\documentclass[conference]{IEEEtran}
\IEEEoverridecommandlockouts
\usepackage{graphicx} %use graph format
\usepackage{epstopdf}
\usepackage{float}
\usepackage{amsmath}
\usepackage{amsthm} 
\newtheorem{myDef}{Definition}

\usepackage{stfloats}
\usepackage{cite}
\usepackage{subfigure}
\usepackage{amsmath,amssymb,amsfonts}
\usepackage{algorithmic}
\usepackage{graphicx}
\usepackage{textcomp}
\usepackage{verbatim}
\usepackage{hyperref}
\hypersetup{hypertex=true,
	colorlinks=true,
	linkcolor=blue,
	anchorcolor=blue,
	citecolor=blue}
\usepackage{xcolor}
\usepackage{algorithmic}%算法
\usepackage{bm}
\usepackage[ruled,vlined]{algorithm2e}
\def\BibTeX{{\rm B\kern-.05em{\sc i\kern-.025em b}\kern-.08em
    T\kern-.1667em\lower.7ex\hbox{E}\kern-.125emX}}

\begin{document}

\title{RFID: Towards Low Latency and Reliable DAG Task Scheduling over Dynamic Vehicular Clouds\\
}

\author{\IEEEauthorblockN{Zhang Liu, \emph{Student Member, IEEE}, Minghui Liwang, \emph{Member, IEEE}, Seyyedali Hosseinalipour, \emph{Member, IEEE}, \\ Huaiyu Dai, \emph{Fellow, IEEE}, Zhibin Gao \emph{Member, IEEE}, Lianfen Huang}
	%Xxxxxx\IEEEauthorrefmark{1}, Xxxxxx\IEEEauthorrefmark{2}\\  
	%\IEEEauthorblockA{\IEEEauthorrefmark{2}\footnotesize 
	%	Dept. of Information and Communication Engineering, Xiamen University, Xiamen, China\\\IEEEauthorrefmark{1}\footnotesize Corresponding author \\ Email: liuzhang@stu.xmu.edu.cn, gaozhibin@xmu.edu.cn}
	
\thanks{
\emph{Zhang Liu (zhangliu@stu.xmu.edu.cn), Minghui Liwang (minghuilw@xmu.edu.cn), Zhibin Gao (gaozhibin@xmu.edu.cn) and Lianfen Huang (lfhuang@xmu.edu.cn) are with the Department of Information and Communication Engineering, School of Informatics, Xiamen University, Fujian, China. S. Hosseinalipour (alipour@buffalo.edu) is with the Department of
Electrical Engineering, University at Buffalo, SUNY, Buffalo, NY, USA. Huaiyu Dai (hdai@ncsu.edu) is with Department of Electrical and Computer Engineering, North Carolina State University, Raleigh, NC, USA. (Corresponding author: Minghui Liwang).} 
}}
%NC 26795 USA
\maketitle

\begin{abstract}
Vehicular cloud (VC) platforms integrate heterogeneous and distributed resources of moving vehicles to offer timely and cost-effective computing services. However, the dynamic nature of VCs (i.e., limited contact duration among vehicles), caused by vehicles' mobility, poses unique challenges to the execution of computation-intensive applications/tasks with directed acyclic graph (DAG) structure, where each task consists of multiple interdependent components (subtasks). In this paper, we study scheduling of DAG tasks over dynamic VCs, where multiple subtasks of a DAG task are dispersed across vehicles and then processed by cooperatively utilizing vehicles' resources. We formulate DAG task scheduling as a 0-1 integer programming, aiming to minimize the overall task completion time, while ensuring a high execution success rate, which turns out to be NP-hard. To tackle the problem, we develop a \underline{r}anking and \underline{f}oresight-\underline{i}ntegrated \underline{d}ynamic scheduling scheme (RFID). RFID consists of (\emph{i}) a \emph{dynamic downward ranking} mechanism that sorts the scheduling priority of different subtasks, while explicitly taking into account for the sequential execution nature of DAG; (\emph{ii}) a \emph{resource scarcity-based priority changing} mechanism that overcomes possible performance degradations caused by the volatility of VC resources; and (\emph{iii}) a \emph{degree-based weighted earliest finish time} mechanism that assigns the subtask with the highest scheduling priority to the vehicle which offers rapid task execution along with reliable transmission links. Our simulation results reveal the effectiveness of our proposed scheme in comparison to benchmark methods.

\end{abstract}


\section{Introduction}
\subsection{Background and Challenges}
Rapid development of Internet of Vehicles (IoV) has led to the emergence of diverse vehicular applications (referred to as \emph{tasks}), e.g., advanced driver assistance system, Netflix streaming, and VTube \cite{b1}. Many of these tasks are computation-intensive and resource-hungry, requiring a massive amount of computation resources to meet their execution requirements, provisioning of which is often beyond the capability of onboard computation equipment of a single vehicle. One approach to process these tasks is to exploit the vehicle-to-infrastructure (V2I) communications and offload them to either remote cloud servers \cite{b2} or edge computing servers \cite{b3}. Nevertheless, V2I connections are not always accessible, e.g., in suburban areas. Also, continuous transfer of data from the vehicles to the cloud servers may incur high traffic congestion on the backhaul links. Furthermore, edge computing servers may not have enough computation resources to satisfy resource demands of a large number of mobile vehicles.

To address the aforementioned limitations, vehicular cloud (VC)\cite{b4,b5,b6} has emerged as an effective computing paradigm, which exploits the dynamic and distributed computation and communication resources of vehicles to provide responsive and cost-effective computing services. Specifically, VC orchestrates the heterogeneous resources of vehicles in geographic proximity and exploits opportunistic vehicle-to-vehicle (V2V) communications to build flexible and yet scaleable topologies for provisioning of computing services.

One of the key advantages of the scalable architecture of VC is its potential to process computation intensive tasks. These tasks are often represented via a directed acyclic graph (DAG)\cite{b15,b16,b17,b18}, where the task is partitioned into interdependent components (referred to as \emph{subtasks}) and the processing intricacies between subtasks is captured via introducing a topological structure among them. 
\begin{figure}[H]
	\centering
	\includegraphics[width=3.5in]{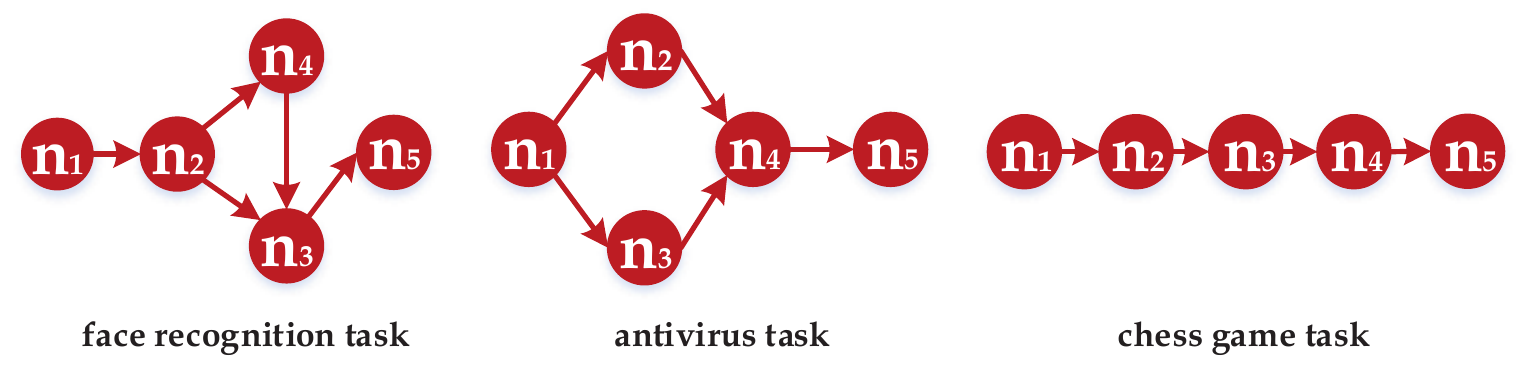}
	\caption{Examples of DAG-represented tasks, and the corresponding interdependence among subtasks\cite{b36}.}
	\label{fig_1}
\end{figure}
Fig. \ref{fig_1} illustrates some examples of DAG tasks, and the corresponding interdependencies among their subtasks\cite{b36}. Take face recognition task as an example, which can be generated by smart buses to trace the contacts of patients who are tested positive during the COVID-19 epidemic\cite{b57}. In a face recognition task, each subtask (vertex) denotes one part of the process of face recognition, while the edges represent the corresponding interdependencies, such as outline and color information. The execution of a DAG task should be conducted in an ordered manner since processing a subtask potentially needs the output data of others (e.g., in the face recognition task in Fig. \ref{fig_1}, the processing of subtask $n_{2}$ relies on the output data of subtask $n_{1}$ and the processing of subtask $n_{3}$ relies on the output data of both $n_{2}$ and $n_{4}$).

Upon execution of a DAG task over a VC, its subtasks can be scheduled and processed by different vehicles in a cooperative manner via V2V communication links. However, there still remains noteworthy issues when scheduling DAG tasks over VCs. First, the heterogeneity of vehicles' processing capabilities (e.g., different computation and communication resources) leads to the non-triviality of determining the scheduling priority of each subtask. Second, dynamic and volatile VC topology leads to time-varying availability of computation resources (e.g., vehicles may arrive at or depart from the VC while a DAG task is being processed), which further adds to the difficulty of subtask ranking and scheduling. Third, given the sequential execution procedure of subtasks within a DAG task, failure in the processing of a single subtask, caused by vehicles' mobility (e.g., upon completion of a subtask, there are no feasible vehicles to execute the subsequent subtasks due to the constrained V2V connections), results in the failure of the entire DAG task. These issues make DAG task scheduling over dynamic VC a challenging problem, which should be carefully investigated while taking into account for the unique characteristics of both the VC and the DAG task structure.

\subsection{Overview and Summary of Contributions}
This paper investigates DAG task scheduling over VC aiming to minimize the overall DAG task completion time (i.e., low latency), while ensuring a high execution success rate (i.e., high reliability). We formulate the DAG task scheduling problem, while explicitly taking into account for: \emph{\romannumeral1)} the \emph{heterogeneity} of computation and communication resources of different vehicles (referred to as resource providers), \emph{\romannumeral2)} \emph{dynamics} and \emph{volatility} of VC's topology, and \emph{\romannumeral3}) the \emph{sequential execution} nature of a DAG task imposed by the interdependencies among its subtasks. We introduce the unified framework of \textbf{r}anking and \textbf{f}oresight-\textbf{i}ntegrated \textbf{d}ynamic scheduling scheme (RFID), which aims to minimize the overall completion time of DAG task, while ensuring a commendable probability of successful task execution.

Our major contributions can be summarized as follows:
\begin{itemize}
	\item  To the best of our knowledge, this paper is among the first to address minimizing DAG task completion time under task execution success rate guarantee over dynamic VC. This is achieved via considering the sequential execution order of subtasks within a DAG task, while capturing the dynamics of VC through a time-varying graph.
	
	\item We formulate the VC-assisted DAG task scheduling problem as an integer programming, aiming to minimize the overall DAG task completion time while ensuring high execution success rate upon considering dynamics and resource heterogeneity of VC, which turns out to be NP-hard.  
	
	\item To tackle the problem, we propose a dynamic scheduling algorithm over VC, called RFID. RFID first recursively determines the scheduling priority of different subtasks according to the assignment of their immediate predecessors. It then selectively changes the scheduling priority of a fraction of subtasks according to the availability of vehicles' resources. Finally, RFID selects the vehicles for subtask assignment based on the vehicles' connectivity and resources.

	\item  We implement RFID over real-world traffic data obtained from the OpenStreetMap\cite{b55}. We further leverage SUMO\cite{b56} simulation platform to simulate the environment and demonstrate the effectiveness of RFID. The simulation results reveal that RFID outperforms benchmark methods in terms of DAG task completion time and execution success rate, while enjoying a relatively low computation complexity.
\end{itemize}

The rest of this paper is organized as follows: Section \uppercase\expandafter{\romannumeral2} discusses related works on task scheduling over different network architectures. In Section \uppercase\expandafter{\romannumeral3}, we present the system model and formulate the VC-assisted DAG task scheduling problem. In Section \uppercase\expandafter{\romannumeral4}, we introduce RFID. Simulation results are presented in Section \uppercase\expandafter{\romannumeral5} and the work in concluded in Section \uppercase\expandafter{\romannumeral6}.

%\begin{figure}[htbp]
%	\centering
%	\subfigure[]{
%		\begin{minipage}[t]{0.5\linewidth}
%			\centering
%			\includegraphics[width=1.5in]{vehicularnetworks.eps}
%			%\caption{fig1}
%		\end{minipage}%
%	}%
%	\subfigure[]{
%		\begin{minipage}[t]{0.5\linewidth}
%			\centering
%			\includegraphics[height=1in,width=2cm]{DAG.eps}
			%\caption{fig2}
%		\end{minipage}%
%	}%
%	\centering
%	\caption{Vehicular cloud and JT.}
%	\label{fig_1}
%\end{figure}

\section{Related Work}
Existing works devoted to task scheduling/offloading over cloud-based networks can be roughly divided into three categories with respect to their task model: \emph{\romannumeral1}) tasks represented by indivisible bit streams with no interdependent subtasks, such as \cite{b7,b8,b9,b10}; \emph{\romannumeral2}) tasks represented via \emph{undirected} graphs considering interdependencies among subtasks, such as \cite{b49,b50,b51,b52}, where the subtasks can be offloaded simultaneously and processed on different servers in parallel; \emph{\romannumeral3}) tasks that are modeled as DAG which further require explicit processing order across their subtasks, e.g., \cite{b40}, \cite{b15,b16,b17,b18}. In the following, we discuss the contributions of these works and highlight the differences between the scenario considered in this paper and prior works.
\subsection{Scheduling of Bit Stream Tasks}
X. Chen \emph{et al.} in \cite{b7} studied computation offloading of bit stream tasks in a mobile edge computing (MEC) network, by formulating a multi-user
computation offloading game, while achieving Nash equilibrium. In \cite{b8}, Y. Mao \emph{et al.} addressed computation offloading of bit stream tasks in MEC with energy harvesting devices via proposing a Lyapunov-based algorithm. S. Bi \emph{et al.} in \cite{b9} studied the computation rate maximization of bit stream tasks in wireless powered MEC through a bi-section search
algorithm and a coordinate descent method. In \cite{b10}, H. Guo \emph{et al.} formulated the MEC offloading problem for bit stream tasks in ultra-dense wireless networks using a two-tiered game-theoretic task offloading scheme. Although the aforementioned works provide useful insights on task scheduling, none of them considers execution of computation intensive tasks, which can be partitioned into multiple subtasks across the computing resources. 
\subsection{Scheduling of Undirected Graph (UG) Tasks}
For UG tasks, J. Ghaderi \emph{et al.} in \cite{b49} proposed a randomized task scheduling algorithm under stochastic task arrival/departure. L. Shi \emph{et al.} in \cite{b50} studied the energy-aware scheduling problem for parallel tasks in cloud by designing a time-efficient scheduling algorithm called TaPRA. In \cite{b51} and \cite{b52}, M. Liwang \emph{et al.} focused on the allocation of computation-intensive graph tasks in IoV and proposed subgraph isomorphism extraction-based low complexity mechanisms. However, UG tasks do not require any specific processing order among their subtasks, where all the subtasks of a UG task can be executed in parallel across the computing resources. As a result, this makes their allocation mechanism different than DAG tasks.
\subsection{Scheduling of DAG Tasks over Static Networks}
DAG task scheduling has been extensively studied in static MEC networks with fully connected servers. H. Topcuoglu \emph{et al.} in \cite{b40} proposed the HEFT algorithm, where each subtask is assigned to the processor that can minimize its corresponding completion time. In \cite{b15}, L. F. Bittencourt \emph{et al.} utilized forward looking attribution to improve the performance of HEFT. M. Aggarwal \emph{et al.} in \cite{b20} developed a genetic algorithm for DAG task scheduling. In \cite{b18}, H. Kanemitsu \emph{et al.} proposed a clustering-based DAG task scheduling algorithm focusing on assigning the subtasks which are located on the critical path to the same processor. G. C. Sih \emph{et al.} in \cite{b16} adopted a dynamic scheduling algorithm, where a global time clock is used to regulate the scheduling process. Recently, in \cite{b21}, Y. Sahni \emph{et al.} introduced a JDOFH algorithm to schedule multiple DAG tasks in a multi-hop collaborative edge computing environment. However, the aforementioned works mainly focus on static networks, ignoring the dynamics and instability of service provisioning, which are significant features of VCs. 
\subsection{Scheduling of DAG Tasks over Dynamic Networks}
There exist few recent works dedicated to DAG task scheduling over dynamic networks. F. Sun \emph{et al.} in \cite{b24} addressed cooperative DAG task scheduling in VC to reduce the overall task completion time via implementing a modified genetic algorithm. In \cite{b25}, H. Liu proposed a policy gradient-based offloading scheme for minimizing the overall DAG task completion time in vehicular networks. In our previous work \cite{b26}, we studied topology-aware DAG task allocation in vehicular networks and proposed a simulated annealing-based task allocation algorithm.

Although the aforementioned works take significant steps toward DAG task scheduling in dynamic networks, they suffer from several limitations, which we aim to address in this work. In particular, in \cite{b24}, scheduling time slots were defined to include the execution and data transmission process, which, however, created redundancy in the task completion time. Also, short V2V communication path life time was not considered in \cite{b25}. Moreover, in our previous work \cite{b26}, the topology of IoV is assumed to remain unchanged during the completion of subtasks. Thus, we are motivated to develop a dynamic scheduling scheme with reasonable computation complexity for DAG task scheduling over VC with volatile V2V links. 
\section{System Model and Problem Formulation}
\begin{figure}[!t]
	\centering
	\includegraphics[width=3.5in]{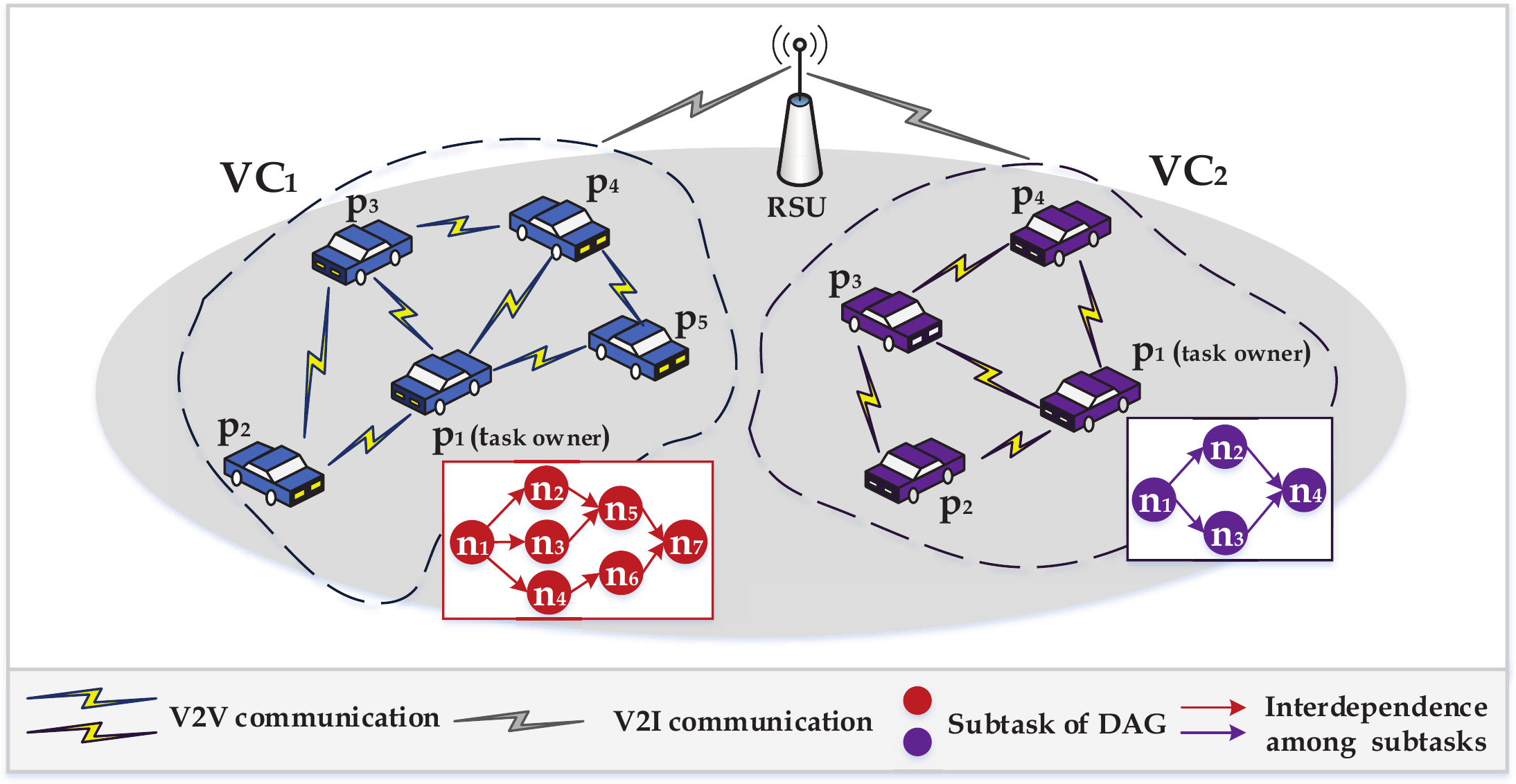}
	\caption{The framework of the proposed VC-assisted DAG task scheduling.}
	\label{fig_2}
\end{figure}
We model each DAG task $A$ as a graph $\mathcal{G}^{\bm{\mathsf{A}}}=\left(\mathcal{V}^{\bm{\mathsf{A}}}, \mathcal{E}^{\bm{\mathsf{A}}}\right)$, where vertex set $\mathcal{V}^{\bm{\mathsf{A}}}=\{n_{1},n_{2}, ...,n_{\left|\mathcal{V}^{\bm{\mathsf{A}}}\right|} \}$ represents the set of subtasks with $\left|\mathcal{V}^{\bm{\mathsf{A}}}\right|$ denoting the number of subtasks in $\bm{\mathsf{A}}$ and $\mathcal{E}^{\bm{\mathsf{A}}}$ denoting the edge set. Each $e^{\bm{\mathsf{A}}}_{n_{i},n_{j}} \in \mathcal{E}^{\bm{\mathsf{A}}}$ is a directed edge describing the corresponding precedence, indicating subtask $n_{i}$ has to be completed before the execution of $n_{j}$ (i.e., the output of $n_{i}$ is used as the input data for $n_{j}$). A subtask without any predecessors represents the \emph{entry} subtask, denoted by $n_{\mathsf{entry}}$, (e.g., subtask $n_{1}$ in Fig. \ref{fig_1}); while a subtask without any successors indicates the exit subtask, denoted by $n_{\mathsf{exit}}$,  (e.g., subtask $n_{5}$ in Fig. \ref{fig_1}). If there is more than one \emph{entry}/\emph{exit} subtask, they are assumed to be connected to a virtual \emph{entry}/\emph{exit} subtask with edges that entail zero data exchange requirements, for analytical simplicity. Fig. \ref{fig_2} depicts a schematic of our system model and some examples, where two VCs are managed by a road side unit (RSU) as a centralized controller\footnote{This paper studies the task scheduling problem via considering one DAG task (generated by task owner) and one VC for analytical simplicity. Our proposed algorithm can also be well applied in networks with multiple VCs and DAGs, e.g., two DAGs in one VC can be seen as a virtual big DAG. Cooperation among VCs and competing for limited resources between multiple DAG tasks are left as our future work.}. Major notations used in this paper are summarized by Table \ref{table1}.

We model the dynamic topology of VC as a time-varying undirected graph $\mathcal{G}^{\bm{\mathsf{VC}}}(t)=\left(\mathcal{V}^{\bm{\mathsf{VC}}}(t), \mathcal{E}^{\bm{\mathsf{VC}}}(t)\right)$. Specifically, the set $\mathcal{V}^{\bm{\mathsf{VC}}}(t)=\{p_{1},p_{2}, ...,p_{\left|\mathcal{V}^{\bm{\mathsf{VC}}}(t)\right|}\}$ contains the task owner $p_{1}$ (i.e., the vehicle who generates the DAG task) and other service vehicles in the network at time $t$; while $\mathcal{E}^{\bm{\mathsf{VC}}}(t)$ represents the corresponding edge set where each $e^{\bm{\mathsf{VC}}}_{p_{m},p_{n}}(t) \in \mathcal{E}^{\bm{\mathsf{VC}}}(t)$ stands for a one-hop two-way V2V link between two vehicles $p_{m}$ and $p_{n}$, where $p_{n}, p_{m} \in \mathcal{V}^{\bm{\mathsf{VC}}}(t)$. The existence of an edge between two vehicles is a result of their corresponding distance as modeled in Section III-B. Vehicles are assumed to have heterogeneous computation capabilities modeled via different local CPU processing speed, denoted by $f_{p_{m}}$ (\emph{cycles/s}) for vehicle $p_{m}$. They also are assumed to execute one subtask at a time \cite{b21}, where multiple subtasks assigned to one vehicle may have to wait for resource release.

\begin{table*}[!t]
	%\scriptsize
	\renewcommand{\arraystretch}{1.3}
	\caption{Major Notations}
	\label{table1}
	\centering
	\begin{tabular}{ll}
		Notations        & \multicolumn{1}{c}{Explanation}                                                                                                                 \\ \hline
		$\mathcal{G}^{\bm{\mathsf{A}}}=\left(\mathcal{V}^{\bm{\mathsf{A}}}, \mathcal{E}^{\bm{\mathsf{A}}}\right)$          & \begin{tabular}[c]{@{}l@{}}DAG task model, where $\mathcal{V}^{\bm{\mathsf{A}}}$ is the set of subtasks and $\mathcal{E}^{\bm{\mathsf{A}}}$ is the set of directed edges\end{tabular}                                 \\
		$\mathcal{G}^{\bm{\mathsf{VC}}}(t)=\left(\mathcal{V}^{\bm{\mathsf{VC}}}(t),\mathcal{E}^{\bm{\mathsf{VC}}}(t)\right)$ & \begin{tabular}[c]{@{}l@{}}VC model, where $\mathcal{V}^{\bm{\mathsf{VC}}}(t)$ contains the vehicles in network at time $t$ and $\mathcal{E}^{\bm{\mathsf{VC}}}(t)$ contains the corresponding one-hop\\ V2V links among vehicles\end{tabular} \\	
		$\mathsf{TT}_{n_{i},n_{j}}(p_{m},p_{n})$ & \begin{tabular}[c]{@{}l@{}}The data transmission time from vehicle $p_{m}$ (assigned to subtask $n_{i}$) to vehicle $p_{n}$ ( processing subtask $n_{j}$)\end{tabular} \\	
		$\mathsf{CT}(n_{i},p_{m})$             & \begin{tabular}[c]{@{}l@{}}The computation time of processing subtask $n_{i}$ on vehicle $p_{m}$\end{tabular}            \\
		$\xi_{n_{i},p_{m}}$              & \begin{tabular}[c]{@{}l@{}}A binary variable indicating the assignment of subtask $n_{i}$ to vehicle $p_{m}$
		\end{tabular}                                 \\
		$\mathsf{st}_{n_{i}}$              & \begin{tabular}[c]{@{}l@{}}The scheduling time of subtask $n_{i}$
		\end{tabular}                                 \\
		$\mathsf{pred}(n_{i})$         & The immediate predecessor set of subtask $n_{i}$                                                                                                     \\
		$\mathsf{succ}(n_{i})$         & The immediate successor set of subtask $n_{i}$                                                                                                        \\
		$\mathsf{RT}(n_{i},p_{m})$          & \begin{tabular}[c]{@{}l@{}}The ready time of processing subtask $n_{i}$ on vehicle $p_{m}$\end{tabular}                              \\
		$\mathsf{EST}(n_{i},p_{m})$         & \begin{tabular}[c]{@{}l@{}}The earliest execution start time of subtask $n_{i}$ on vehicle $p_{m}$\end{tabular}                                         \\
		$\mathsf{EFT}(n_{i},p_{m})$         & \begin{tabular}[c]{@{}l@{}}The earliest execution finish time of subtask $n_{i}$ on vehicle $p_{m}$\end{tabular}                                         \\
		$\mathsf{AFT}(n_{i})$         & \begin{tabular}[c]{@{}l@{}}The actual finish time of subtask $n_{i}$ when $n_{i}$ is practically processed on a specific vehicle \end{tabular}                                         \\
		$\mathcal{C}^{\bm{\mathsf{VC}}}(n_{i})$ & \begin{tabular}[c]{@{}l@{}}The candidate vehicle set for scheduling subtask $n_{i}$  \end{tabular} \\
		$\bm{n_{\mathsf{ready}}}$ & \begin{tabular}[c]{@{}l@{}}The time-varying ready subtask set  \end{tabular} \\
		$\mathcal{D}^{\bm{\mathsf{VC}}}(n_{i},p_{m})$ & \begin{tabular}[c]{@{}l@{}}The degree-based vehicle set when subtask $n_{i}$ is processed on vehicle $p_{m}$ \end{tabular} \\
		$\mathsf{EFT}^{\mathsf{W}}(n_{i},p_{m})$         & \begin{tabular}[c]{@{}l@{}}The weighted earliest finish time of subtask $n_{i}$ on vehicle $p_{m}$\end{tabular}                                         \\
		$n_{i},n_{j}$             & Indices used to represent subtasks                                                                                                                 \\
		$p_{m},p_{n}$          & Indices used to represent vehicles    
		\\ \hline                                                                                                           
	\end{tabular}
\end{table*}
\subsection{Communication Model}
Considering transmitting the processing results of subtask $n_{i}$ (executed on vehicle $p_{m}$) to vehicle $p_{n}$ which is assigned to execute subtask $n_{j}$, where $e^{\bm{\mathsf{A}}}_{n_{i},n_{j}} \in \mathcal{E}^{\bm{\mathsf{A}}}$, a dual-slope (power-law) model\cite{b58} is leveraged to formulate the propagation loss of the underlying V2V communication link in dB, which is considered to be full-duplex as follows:
\begin{equation}
\mathsf{PL}\left(d_{p_{m}, p_{n}}(t)\right)=L_{b}+\left\{\begin{array}{l}
	10 \eta_{1} \log \left(d_{p_{m}, p_{n}}\left(t\right)\right)+
	\mathsf{PL}(d_{0}), \\ \quad\quad\quad\quad \text { if } 1 \leq d_{p_{m}, p_{n}}(t) \leq d_{\mathsf{brk}}\\
	10 (\eta_{1}-\eta_{2}) \log \left(d_{\mathsf{brk}}\right)+ \\
	10 \eta_{2} \log \left(d_{p_{m}, p_{n}}\left(t\right)\right)+\mathsf{PL}(d_{0}), \\\quad\quad\quad\quad\quad \text { if } d_{p_{m}, p_{n}}(t)>d_{\mathsf{brk}}
\end{array}\right.\label {eq_1}
\end{equation}
where $L_{b}$ is a basic transmission-loss parameter that depends on the frequency and the antenna height. $d_{p_{m}, p_{n}}(t)$ indicates the Euclidean distance between vehicle $p_{m}$ and $p_{n}$ at time $t$, $d_{0}=1$ (\emph{m}) is the reference distance, $\eta_{1}=2$ and $\eta_{2} \in [2,7]$ denote the slopes of the best-fit line before and after distance $d_{\mathsf{brk}}$, respectively, and $d_{\mathsf{brk}}$ indicates the breakpoint distance given by
\begin{equation}
	d_{\mathsf{brk}}=\frac{4 h_{t} h_{r}}{\delta}-\frac{\lambda}{4}, \label {eq_2}
\end{equation}
where $h_{t}$ and $h_{r}$ are the height of transmitter (i.e., the antenna on vehicle $p_{m}$) and the receiver (i.e., the antenna on vehicle $p_{n}$), $\delta$ represents the signal power fluctuations due to surrounding objects, and $\lambda$ denotes the wavelength\cite{b61}.

Let the binary indicator variable $\xi_{n_{i},p_{m}}$ describe the subtask assignment: $\xi_{n_{i},p_{m}}=1$ if subtask $n_{i}$ is scheduled on vehicle $p_{m}$; otherwise $\xi_{n_{i},p_{m}}=0$. The transmission time (\emph{s}) associated with data transmission over the edge $e^{\bm{\mathsf{A}}}_{n_{i},n_{j}} \in \mathcal{E}^{\bm{\mathsf{A}}}$, when $\xi_{n_{i},p_{m}} \times \xi_{n_{j},p_{n}}=1$ at time $t$ is given by
\begin{equation}
	\mathsf{TT}_{n_{i}, n_{j}}\left(p_{m}, p_{n}\right)=\left\{\begin{array}{l}
		c_{n_{i}, n_{j}}\times{\Gamma\left(\mathsf{PL}\left(d_{p_{m}, p_{n}}\left(t\right)\right)\right)}, p_{m} \neq p_{n} \\
		0, \quad\quad\quad\quad\quad\quad\quad\quad\quad\quad\ \text { otherwise }
	\end{array}\right.\label {eq_3}
\end{equation}
where $c_{n_{i},n_{j}}$ denotes the size (\emph{bit}) of output data of subtask $n_{i}$, which needs to be transferred to its dependent task $n_{j}$, and $\Gamma( )$ is a monotone increasing function expressed in\cite{b59}. Due to vehicles' mobility and their limited contact durations, this paper only considers one-hop data transmission between the vehicles \cite{b4,b5,b6}.

\subsection{V2V Contact Model}
A contact event between vehicles $p_{m}$ and $p_{n}$ can happen when $d_{p_{m}, p_{n}}(t) \leq R$ at time $t$, where \emph{R} represents the vehicular communication radius. The contact duration between $p_{m}$ and $p_{n}$ is assumed to obey an exponential distribution \cite{b60}
with parameter $\mu_{p_{m},p_{n}}$. Correspondingly, the probability that the residual contact duration denoted by $\left|e_{p_{m}, p_{n}}^{\bm{\mathsf{VC}}}(t)\right|$ between vehicle $p_{m}$ and $p_{n}$ at time $t$ is larger than $T$ is given by
\begin{equation}
	\operatorname{Pr}\left(\left|e_{p_{m}, p_{n}}^{\bm{\mathsf{VC}}}(t)\right|>T\right)=\exp \left(-T \mu_{p_{m}, p_{n}}\right).\label {eq_4}
\end{equation}
According to (\ref{eq_4}), a lower value of the required contact duration $T$ for data transmission can be provided with a higher assurance.  

\subsection{Computation Model}
Let $\mathsf{EST}(n_{i},p_{m})$ and $\mathsf{EFT}(n_{i},p_{m})$ denote the earliest execution start time of subtask $n_{i}$ on $p_{m}$, and the earliest execution finish time of subtask $n_{i}$ on vehicle $p_{m}$, respectively. 

For the entry subtask, $n_{\mathsf{entry}}$ is assumed to be executed on the task owner $p_{1}$, and thus $\mathsf{EST}(n_{\mathsf{entry}}, p_{1})=0$. 

Next, we provide key definitions used in developing our methodology.
\begin{myDef}
	\textbf{(Ready time).} The ready time $\mathsf{RT}\left(n_{i}, p_{m}\right)$ is time when all the immediate predecessor subtasks of $n_{i}$ have been completed, while the required input data for processing $n_{i}$ has arrived at vehicle $p_{m}$, which is given by
	\begin{equation}
		\mathsf{RT}\left(n_{i}, p_{m}\right)=\max _{\substack{n_{j} \in \mathsf{p r e d}\left(n_{i}\right), \\ \xi_{n_{j}, p_{n}}=1}}\left\{\mathsf{AFT}\left(n_{j}\right)+\mathsf{TT}_{n_{j}, n_{i}}\left(p_{n}, p_{m}\right)\right\},\label {eq_5}
	\end{equation}
	where $\mathsf{p r e d}\left(n_{i}\right)$ is the set of immediate predecessor subtasks of $n_{i}$ and $\mathsf{AFT}(n_{j})$ is the actual finish time of $n_{j}$ when it is practically scheduled on a specific vehicle:
	\begin{equation}
        \mathsf{AFT}\left(n_{j}\right)=\mathsf{EFT}\left(n_{j}, p_{m}\right) \text {, where } \xi_{n_{j}, p_{m}}=1.\label {eq_6}
	\end{equation}
\end{myDef}
\begin{myDef}
	\textbf{(Scheduling time).} The scheduling time of subtask $n_{i}$, i.e., $\mathsf{st}_{n_{i}}$ is the earliest time to allocate $n_{i}$ to a vehicle, which is given by 
	\begin{equation}
		\mathsf{st}_{n_{i}}=\max _{n_{j} \in \mathsf{pred}\left(n_{i}\right)}\left\{\mathsf{AFT}\left(n_{j}\right)\right\},\label {eq_7}
	\end{equation}
\end{myDef}

Where, $\mathsf{pred}(n_{i})$ is the set of immediate predecessors of subtask $n_{i}$. Using Definition 1, for the other subtasks involved in the DAG, the values of EFT and EST can be computed recursively, starting from the \emph{entry} subtask as follows:
	\begin{equation} 		
	\mathsf{EST}(n_{i}, p_{m})=\max \left\{ \mathsf{Avail}(n_{i},p_{m}), \mathsf{RT}(n_{i}, p_{m})\right\},\label {eq_8}
	\end{equation} 
	\begin{equation}
	\mathsf{EFT}(n_{i}, p_{m})=\mathsf{CT}(n_{i},p_{m})+\mathsf{EST}(n_{i}, p_{m}),\label {eq_9}
	\end{equation}
where $\mathsf{Avail}(n_{i},p_{m})$ represents the time, in which vehicle $p_{m}$ completes its last assigned subtask prior to the execution of $n_{i}$ and it is ready to process a new subtask. Also, $\mathsf{CT}(n_{i},p_{m})$ denotes the computation time of processing subtask $n_{i}$ on vehicle $p_{m}$, which is given by
	\begin{equation}
	\mathsf{CT}(n_{i},p_{m})=w_{n_{i}}/f_{p_{m}},\label {eq_10}
	\end{equation}
where $w_{n_{i}}$ represents the required computing workload (\emph{cpu clock cycles}) of subtask $n_{i}$.

Since there can be multiple \emph{exit} subtasks, after all the subtasks in a DAG are scheduled,  the actual finish time of the \emph{exit} subtask $n_\mathsf{exit}$ is defined as the overall DAG task completion time, which is given by 
	\begin{equation}
	\mathsf{OTC}=\mathsf{AFT}(n_{\mathsf{exit}}).\label {eq_11}
	\end{equation}

The main goal of DAG task scheduling is to determine the assignment of each subtask to the appropriate vehicle such that the overall DAG task completion time can be minimized.

\subsection{Problem Formulation}
We formulate DAG task scheduling over dynamic VC as the following optimization problem:
\begin{flalign}
	\bm{\mathcal{P:}}\quad \underset{\{\xi_{n_{i}, p_{m}}\}, \ n_{i} \in \mathcal{V}^{\bm{\mathsf{A}}},\ p_{m} \in \mathcal{V}^{\bm{\mathsf{VC}}}\left(\mathsf{st}_{n_{i}}\right)}{\arg \min } \ \mathsf{OTC} \label {eq_12}              
\end{flalign}
\begin{equation}
	\begin{aligned}
		&\text{s.t. } \ \textbf{C1:} \sum_{p_{m} \in \mathcal{V}^{\bm{\mathsf{VC}}}(\mathsf{st}_{n_{i}})} \xi_{n_{i}, p_{m}}=1, \nonumber \\
	 &\textbf{C2:}\ \mathsf{EST}\left(n_{i},p_{m}\right) \geq \max _{n_{j} \in \mathsf
	 	{pred}\left(n_{i}\right)}\left\{\mathsf{A F T}\left(n_{j}\right)\right\}, \ \xi_{n_{i}, p_{m}}=1,\nonumber \\
	 &\textbf{C3:}\ \exp \left(-\mathsf{TT}_{n_{j}, n_{i}}\left(p_{n}, p_{m}\right)\mu_{p_{n}, p_{m}} \right) \geq \theta,\nonumber \\
	 &\forall n_{j} \in \mathsf{pred} \left(n_{i}\right), and\ \xi_{n_{i},p_{m}} \times \xi_{n_{j},p_{n}}=1, \nonumber\\
	 &\textbf{C4:}\ \xi_{n_{i}, p_{m}} \in\{0,1\}.\nonumber
	\end{aligned}
\end{equation}
In problem $\bm{\mathcal{P}}$, \textbf{C1} guarantees that each subtask $n_{i}$ can be assigned to only one vehicle, \textbf{C2} represents that execution of a subtask can not start until all its predecessor subtasks are completed based on the sequential execution property of DAG task. Considering the volatility of V2V links modeled in Section III-B, \textbf{C3} ensures that vehicle $p_{m}$, scheduled to execute subtask $n_{i}$, can successfully receive input data of $n_{i}$, where $\theta$ in \textbf{C3} is the predefined quality of service factor. Besides, \textbf{C4} restricts that the scheduling variables are binary.

$\bm{\mathcal{P}}$ is a 0-1 integer programming problem which is NP-hard. This makes finding time-efficient algorithms to solve the problem difficult, especially in large-scale dynamic networks. Also, solving $\bm{\mathcal{P}}$ requires determining the scheduling priority of each subtask to preserve the sequential processing requirements imposed by the DAG structure in volatile VC environment, which is challenging since during the execution of subtasks the links among the vehicles may begin to form or vanish. Furthermore, since the completion of subtask $n_{exit}$ is considered as the objective function, if the execution of any intermediate subtasks fails (e.g., there are no feasible vehicles for processing a subtask due to the constrained V2V connections, i.e., \textbf{C3} can not be satisfied), the execution of entire DAG task will encounter failure. Motivated by these challenges, we develop RFID, which aims to reduce the completion time of task execution, while providing reliability assurance. RFID will enjoy a polynomial time complexity, which makes it suitable for implementation over large-scale dynamic VC.

\section{Ranking and Foresight-Integrated Dynamic (RFID) Task Scheduling}
\begin{figure}[!t]
	\centering
	\includegraphics[width=3.5in]{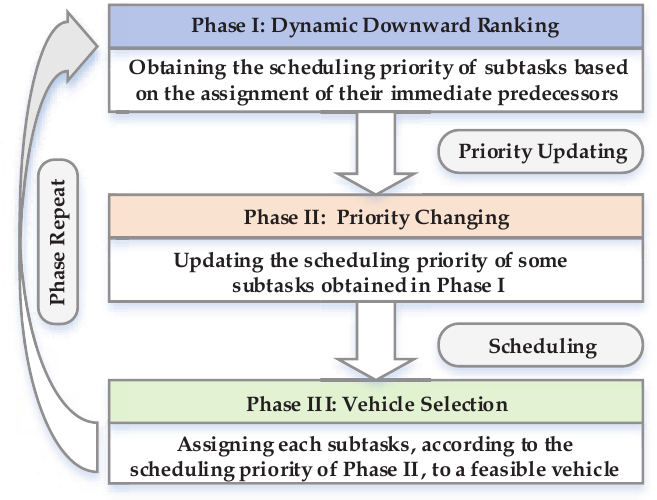}
	\caption{A flow chart of RFID and the inter-relationship between different phases.}
	\label{fig_3}
\end{figure}

To tackle $\bm{\mathcal{P}}$, we develop RFID which is a unified DAG task scheduling methodology over dynamic networks. RFID conducts task scheduling through three phases. In phase I, a \emph{dynamic downward ranking} mechanism is designed to sort the scheduling priority of different subtasks over dynamic VC, via considering the assignment of their immediate predecessor subtasks, to capture DAG task's sequential execution nature. In phase II, a \emph{resource scarcity-based priority changing} mechanism is developed to overcome possible performance degradations caused by the volatility of VC resources via modifying the scheduling priority of a fraction of subtasks obtained in phase I. Finally, in phase III, a \emph{degree-based weighted earliest finish time} mechanism is introduced to assign the subtask with the highest scheduling priority to the vehicle which offers rapid task execution, which also possesses reliable V2V links. A flow chart of RFID, and the inter-relationship between the above-mentioned three phases is shown by Fig. \ref{fig_3}. Before discussing RFID, we first present the definition of \emph{candidate set}, which is considered as a significant part to develop RFID.

\begin{myDef}
	\textbf{(Candidate set).} The candidate set $\mathcal{C}^{\bm{\mathsf{VC}}}\left(n_{i}\right)$ of $n_{i}$ contains the vehicles that can receive the output data of all the predecessors of $n_{i}$, as given by		
\begin{equation}
	\begin{aligned}
	\mathcal{C}^{\bm{\mathsf{VC}}}\left(n_{i}\right) \triangleq	
	\{p_{m}\!:\!\exp \left(-\mathsf{TT}_{n_{j}, n_{i}}\left(p_{n}, p_{m}\right) \mu_{p_{n}, p_{m}} \right) \!\geq\! \theta, \\
	p_{m}\in \mathcal{V}^{\bm{\mathsf{VC}}}(st_{n_{i}}),\ n_{j} \in \mathsf{pred} \left(n_{i}\right), \ \xi_{n_{j},p_{n}}=1 \}, \label {eq_13}
	\end{aligned}
\end{equation}
where $\mathsf{pred}(n_{i})$ is the set of immediate predecessors of $n_{i}$, and $\theta$ represents the predefined quality of service factor, as defined in \textbf{C3}.
\end{myDef}

In the following, details of the three phases of RFID are discussed. In each phase, we first highlight the shortcomings of the current state-of-the-art methods, and then, develop our methodology.
\subsection{Phase I: Dynamic Downward Ranking}
The major goal of this phase is to determine the scheduling priority of subtasks, based on a metric called \emph{ranking}, where the subtask with a lower rank is considered to have a higher scheduling priority.

\subsubsection{Motivation}
Existing methods such as \cite{b15,b16,b17},\cite{b21} focus on subtask ranking in static environments, e.g., MEC networks, with multiple fully-connected processors. They determine the scheduling of subtasks based on \emph{downward ranking}, which can be obtained recursively by traversing the DAG downward (i.e., starting from $n_{\mathsf{entry}}$ to $n_{\mathsf{exit}}$). For $n_{\mathsf{entry}}$, it is assumed that $\mathsf{rank}(n_{\mathsf{entry}})=0$, while for the other subtasks, we have the following (\ref{eq_14})
\begin{equation}
	\begin{split}
		\mathsf{rank}\left(n_{i}\right)=\!\!\!\max _{n_{j} \in \mathsf{pred}\left(n_{i}\right)}\{\mathsf{rank}\left(n_{j}\right)+\overline{\mathsf{CT}}_{n_{i}}+\overline{\mathsf{TT}}_{n_{j}, n_{i}}\}, \label {eq_14}	
	\end{split}		
\end{equation}
where $\mathsf{pred}(n_{i})$ denotes the set of immediate predecessors of subtask $n_{i}$, $\overline{\mathsf{CT}}_{n_{i}}=\sum_{m=1}^{q} \mathsf{CT}\left(n_{i}, p_{m}\right) / q$ is the average computation time of subtask $n_{i}$ across $q$ static processors, and $\overline{\mathsf{TT}}_{n_{j}, n_{i}}=c_{n_{i}, n_{j}} / \bar{B}$ is the average data transmission time associated with edge $e^{\bm{\mathsf{A}}}_{n_{j},n_{i}}$. $\bar{B}$ denotes the average transmission rate among $q$ static processors. 

\begin{figure}[H]
	\centering
	\includegraphics[width=3.5in]{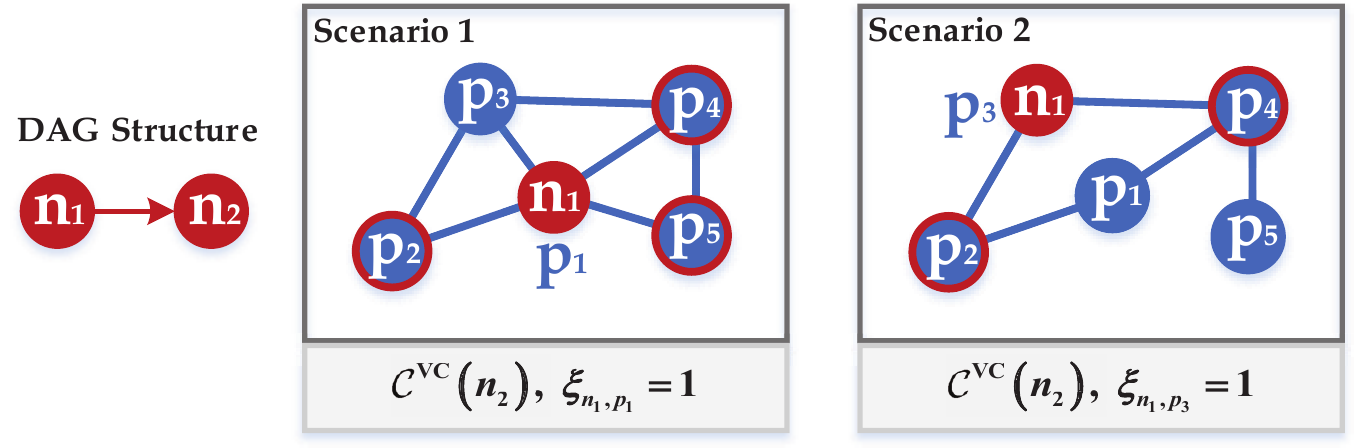}
	\caption{An example of showing the necessity of using dynamic ranking to determine the priority of each subtask over VC.}
	\label{fig_4}
\end{figure}

Since conventional \emph{downward ranking} mainly focuses on static networks (e.g., $q$ and $\bar{B}$ are constants), which, however, does not hold in dynamic VC. To further reveal that existing strategies are difficult to be implemented in our problem setting, we depict a simple example as shown in Fig. \ref{fig_4}, where two vehicles are connected by an edge when the distance between them are smaller than the vehicular communication radius. Due to the mobility and the heterogeneous resources of vehicles, the corresponding candidates for scheduling $n_{2}$ can be different according to different assignments of $n_{1}$. For example, in Fig. \ref{fig_4}, when $n_{1}$ is processed on $p_{1}$, we have $\mathcal{C}^{\bm{\mathsf{VC}}}\left(n_{2}\right)=\{p_{1},p_{2},p_{4},p_{5}\}$ (although there exists an edge between $p_{1}$ and $p_{3}$, $p_{3}$ is excluded in the candidate set due to $\text{Pr}(\left|e_{p_{1}, p_{3}}^{\bm{\mathsf{VC}}}(t)\right|>\mathsf{TT}_{n_{1},n_{2}}(p_{1},p_{3}))<\theta$). However, when $n_{1}$ is processed on $p_{3}$, we have $\mathcal{C}^{\bm{\mathsf{VC}}}\left(n_{2}\right)=\{p_{2},p_{3},p_{4}\}$. Note that the topology of VC changes in the two scenarios considered in Fig. \ref{fig_4}, since the processing time of $n_{1}$ on $p_{1}$ might be different with that on $p_{3}$, which thus results in different values of $q$ and $\bar{B}$. 

As a result, the set of candidate vehicles for processing $n_{i}$, can be impacted by different assignments of $n_{j}$ ($n_{j} \in \mathsf{pred} \left(n_{i}\right)$) due to dynamics, which should be carefully considered during the subtask ranking mechanism design.
\subsubsection{Proposed Dynamic Downward Ranking Mechanism}
To address the aforementioned challenges, we propose a \emph{dynamic downward ranking} mechanism, which is tightly coupled with the vehicle selection method described in phase III. We first provide the definition of a \emph{ready subtask}.
\begin{myDef}
	\textbf{(Ready subtask).} The set of ready subtask\footnote{Although $\bm{n_{\mathsf{ready}}}$ is time varying, for notational simplicity, we consider $\bm{n_{\mathsf{ready}}}$ only changes during the algorithm execution.} $\bm{n_{\mathsf{ready}}}$ contains the subtasks whose immediate predecessors have already been scheduled, and the corresponding execution order is not bounded by a precedence constraint, i.e, $\forall n_{i}, n_{j} \in \bm{n_{\mathsf{ready}}}, e^{\bm{\mathsf{A}}}_{n_{i},n_{j}} \notin \mathcal{E}^{\bm{\mathsf{A}}}$.
\end{myDef}

We next model the \emph{dynamic downward ranking} $\mathsf{Rank}^{\mathsf{D}}$ of each subtask $n_{i} \in \bm{n_{\mathsf{ready}}}$, by leveraging available resources and topological information of VC. In particular, the value of $\mathsf{Rank}^{\mathsf{D}}(n_{i})$ can be different according to different assignments of its immediate predecessor subtasks. 

For the entry subtask, we have $\mathsf{Rank}^{\mathsf{D}}(n_{\mathsf{entry}})=0$. After $n_{\mathsf{entry}}$ is scheduled on task owner (according to our basic assumption in previous sections), we determine the \emph{ranking} of the existing subtask $n_{i} \in \bm{n_{\mathsf{ready}}}$ and subsequently assign them to appropriate vehicles (based on the vehicle selection method, which will be described in phase III). To this end, we compute the value of \emph{dynamic downward ranking} as the following (\ref{eq_15}),
\begin{equation}
	\begin{split}
		\mathsf{Rank}^{\mathsf{D}}\left(n_{i}\right)= \!\!\!\max _{n_{j} \in \mathsf{pred}\left(n_{i}\right)}\{\mathsf{Rank}^{\mathsf{D}}\left(n_{j}\right)+\overline{\mathsf{CT}}^{\mathsf{D}}_{n_{i}}+\overline{\mathsf{TT}}^{\mathsf{D}}_{n_{j}, n_{i}}\}, 
		\\ n_{i} \in \bm{n_{\mathsf{ready}}},  \label {eq_15}	
	\end{split}		
\end{equation}
where $\mathsf{pred}(n_{i})$ is the set of immediate predecessors of subtask $n_{i}$, $\overline{\mathsf{CT}}^{\mathsf{D}}_{n_{i}}$ is the dynamic average computation time, impacted by the assignment of $n_{j} \in \mathsf{pred}(n_{i})$, as given by
\begin{equation}
	\begin{aligned}
	\overline{\mathsf{CT}}^{\mathsf{D}}_{n_{i}}=\frac{ \sum_{p_{m} \in \mathcal{C}^{\bm{\mathsf{VC}}}\left(n_{i}\right)} \mathsf{CT}\left(n_{i}, p_{m}\right)}{\left|\mathcal{C}^{\bm{\mathsf{VC}}}\left(n_{i}\right)\right|}, \ n_{i} \in \bm{n_{\mathsf{ready}}}. \label {eq_16}
	\end{aligned}
\end{equation}
Also in (\ref{eq_15}), $\overline{\mathsf{TT}}^{\mathsf{D}}_{n_{j},n_{i}}$ represents the dynamic average data transmission time, affected by the assignment of $n_{j} \in \mathsf{pred}(n_{i})$, which is shown by
\begin{equation}
	\begin{aligned}
	\overline{\mathsf{TT}}^{\mathsf{D}}_{n_{j},n_{i}}=\frac{ \sum_{p_{m} \in \mathcal{C}^{\bm{\mathsf{VC}}}\left(n_{i}\right)} \mathsf{TT}_{n_{j}, n_{i}}(p_{n},p_{m})}{\left|\mathcal{C}^{\bm{\mathsf{VC}}}\left(n_{i} \right)\right|}, \ n_{i} \in \bm{n_{\mathsf{ready}}}.\label {eq_17}	
	\end{aligned}	
\end{equation}

As a result, different from static network environments, our \emph{dynamic downward ranking} mechanism identifies the suitability of subtask assignment to different vehicles, which will interactively work with the vehicle selection method described in phase III, until the exit subtask has been scheduled.
\subsection{Phase II: Resource Scarcity-based Priority Changing}
The goal of this phase is to adjust the scheduling priority of a fraction of subtasks obtained in phase I aiming to overcome possible performance degradations caused by the volatility of VC resources.
\subsubsection{Motivation}
\begin{figure}[H]
	\centering
	\includegraphics[width=3.5in]{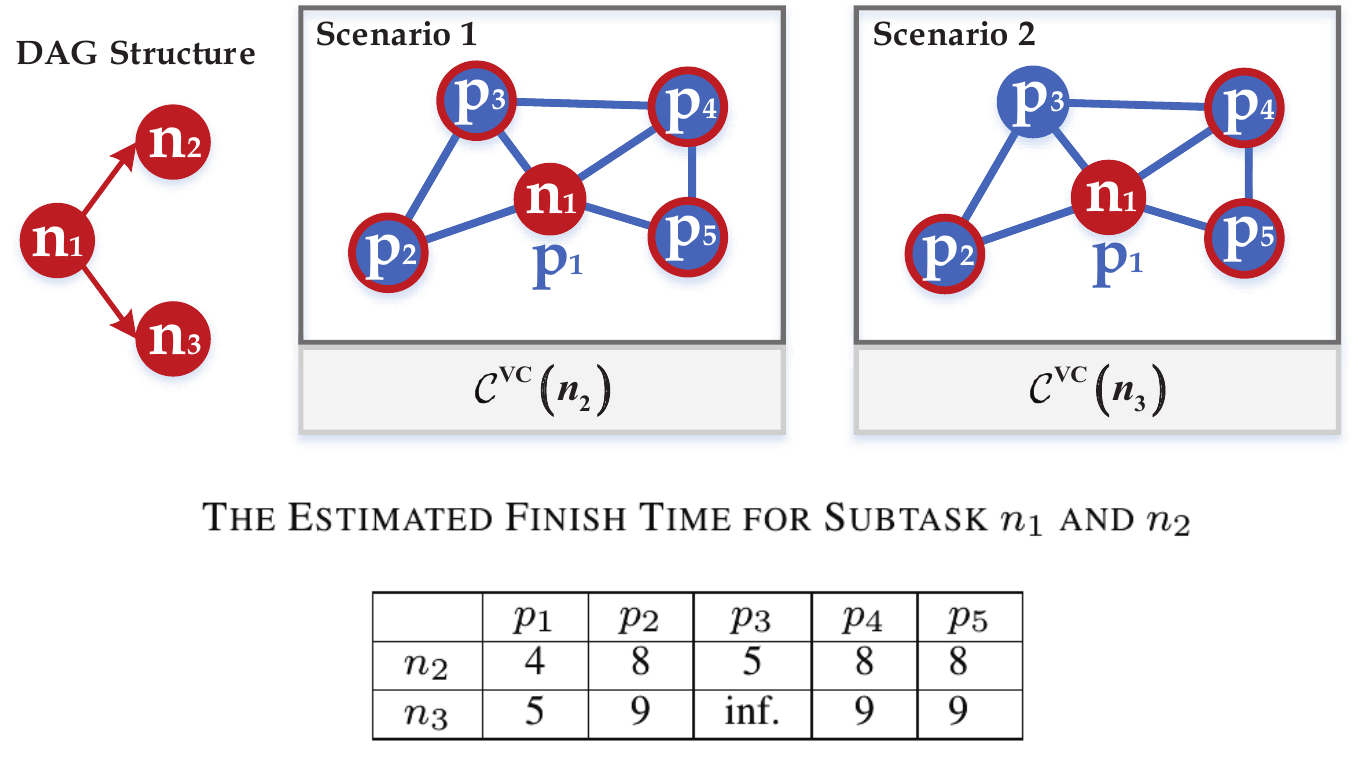}
	\caption{An example showing the necessity of considering resource scarcity in determining the scheduling priority of some specific subtasks over VC.}
	\label{fig_5}
\end{figure}

Different from static computing environments with stable fully-connected computing servers\cite{b15,b16,b17}, dynamics and instability of VC resources can lead to resource scarcity, which leaves heave impacts on the execution of dependent subtasks. To illustrate this, we depict a simple example in Fig. \ref{fig_5}, where after the assignment of $n_{1}$, subtasks $n_{2}$ and $n_{3}$ become \emph{ready subtask} according to Definition 4. Suppose that  $\mathsf{Rank}^{\mathsf{D}}(n_{2})<\mathsf{Rank}^{\mathsf{D}}(n_{3})$, subtask $n_{2}$ will be firstly scheduled to its most preferred vehicle $p_{1}$. However, this case can incur a large completion time increasing, if $n_{3}$ is not assigned to $p_{1}$, because $p_{3}$ can execute $n_{2}$ almost as rapidly. Conversely, since the scale of candidate set for processing $n_{3}$ is smaller (e.g., $p_{3}$ is infeasible for processing $n_{3}$ in Fig. \ref{fig_5}), this can raise the completion time when $n_{3}$ is not assigned to $p_{1}$, as $p_{1}$ is the only vehicle which can process $n_{1}$ quickly. Similar situations can be incurred when more than one \emph{ready subtask}s with different transmission requirements compete for a same vehicle, over dynamic VC with volatile resources.

\subsubsection{Proposed Resource Scarcity-based Priority Changing Mechanism}
Motivated by the aforementioned example, we consider dynamic resource availability caused by vehicles' mobility and address the resource scarcity problem via introducing a new metric denoted by $\mathsf{CTI}(n_{i})$, which measures the completion time increment imposed by the case where subtask $n_{i}$ is not scheduled to its most preferred vehicle, defined as
\begin{equation}
	\begin{split}
		\mathsf{CTI}(n_{i})=\mathsf{E F T}\left(n_{i}, p_{m}^{*}\right)-\min _{p_{m} \neq p_{m}^{*}}\left\{\mathsf{E F T}\left(n_{i}, p_{m}\right)\right\}, \\ p_{m},\ p^{*}_{m} \in \mathcal{C}^{\bm{\mathsf{VC}}}(n_{i}),\label {eq_18}
	\end{split}	
\end{equation} 
where $p_{m}^{*}$ represents the most preferred vehicle with a minimum value of EFT. Specifically, the minimization term in (\ref{eq_18}) captures the earliest finish time of having $n_{i}$ to be executed on the second preferred vehicle.

Consequently, we define the resource-scarcity-based dynamic downward ranking $\mathsf{RSRank}^{\mathsf{D}}(n_{i})$, to determine the scheduling priority of each subtask $n_{i} \in \bm{n_{\mathsf{ready}}}$ as follows: 
\begin{equation}
	\begin{split}
	\mathsf{RSRank}^{\mathsf{D}}(n_{i}) = \mathsf{Rank}^{\mathsf{D}}(n_{i})-\mathsf{CTI}(n_{i}), \ n_{i} \in \bm{n_{\mathsf{ready}}}. \label {eq_19}
	\end{split}	
\end{equation} 
As a result, considering the extra completion time increment in assigning each \emph{ready subtask} to the second preferred vehicle (in terms of completion time), can generally bring a change of scheduling priority. For example, in Fig. \ref{fig_5}, $\mathsf{Rank}^{\mathsf{D}}(n_{2})<\mathsf{Rank}^{\mathsf{D}}(n_{3})$ while $\mathsf{RSRank}^{\mathsf{D}}(n_{3})<\mathsf{RSRank}^{\mathsf{D}}(n_{2})$, and thus RFID schedules $n_{3}$ earlier to avoid a large completion time increment. 

\subsection{Phase III: Degree-based Weighted EFT for Vehicle Selection}
Based on the scheduling priority obtained in Phase II, the goal of this phase is to assign each subtask to the vehicle which can rapidly execute it and have reliable transmission V2V links, which is necessary to transmit the data required for the processing of its successor subtasks.

\subsubsection{Motivation}
\begin{figure}[H]
	\centering
	\includegraphics[width=3.5in]{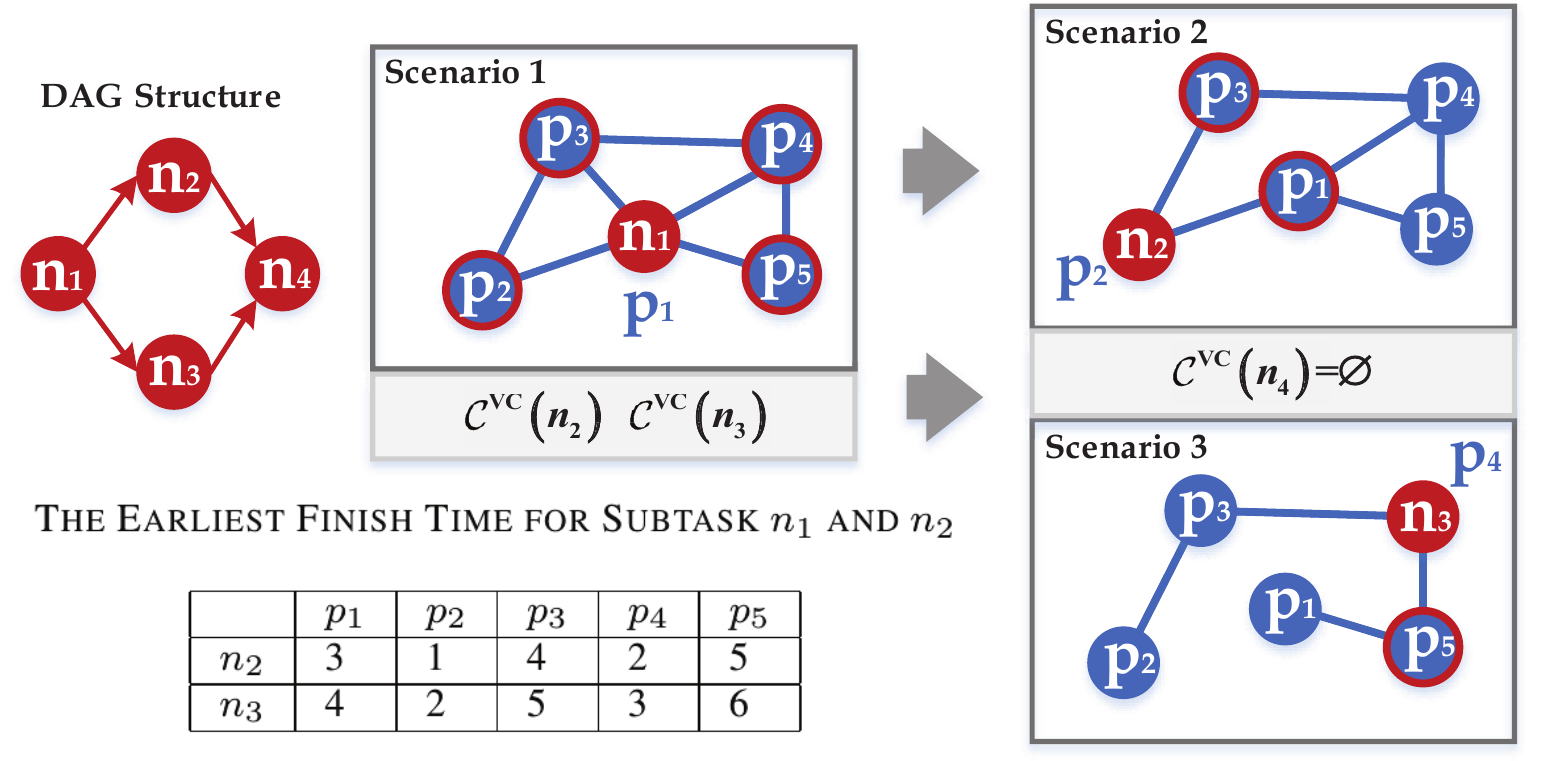}
	\caption{An example of using HEFT to schedule DAG task over dynamic VC.}
	\label{fig_6}
\end{figure}

%\begin{table}
%	\centering
%	\caption{The Earliest Finish Time for Subtask $n_{1}$ and $n_{2}$}
%	\begin{tabular}{|l|l|l|l|l|l|} 
%		\hline
%		& $p_{1}$ & $p_{2}$ & $p_{3}$ & $p_{4}$ & $p_{5}$  \\ 
%		\hline
%		$n_{2}$ & 3  & 1  & 4  & 2  & 5   \\ 
%		\hline
%		$n_{3}$ & 4  & 2  & 5  & 3  & 6   \\
%		\hline
%	\end{tabular}
%\end{table}
Heterogeneous earliest finish time algorithm (HEFT) \cite{b40}, which assigns $n_{i}$ to $p_{m}$ to minimize $\mathsf{EFT}(n_{i},p_{m})$ is widely used in static networks. However, this algorithm is not practical in dynamic VC. We demonstrate this via a simple example shown in Fig. \ref{fig_6}, where the topology of VC changes across three scenarios, since the processing and transmission time of different vehicles can be different, and we have the scheduling order of each subtask as $n_{1} \rightarrow n_{2} \rightarrow n_{3} \rightarrow n_{4}$. After $n_{1}$ ($n_{\mathsf{entry}}$) is scheduled on $p_{1}$ (task owner),  $n_{2}$ and $n_{3}$ are executed on $p_{2}$ and $p_{4}$, respectively, according to the corresponding EFT associated with different vehicles, under HEFT. However, it can be seen that the execution of subtask $n_{4}$ fails since $\mathcal{C}^{\bm{\mathsf{VC}}}(n_{4})=\varnothing$. Because $p_{1}$ and $p_{3}$ are feasible to transmit the output data of $n_{2}$, and $p_{5}$ is applicable to transmit the output data of $n_{3}$ (although $p_{4}$ and $p_{3}$ are connected, $p_{3}$ is infeasible since $\text{Pr}(\left|e_{p_{4}, p_{3}}^{\bm{\mathsf{VC}}}(t)\right|>\mathsf{TT}_{n_{3},n_{4}}(p_{4},p_{3}))<\theta$), while the execution of $n_{4}$ requires the output data from all its immediate predecessor subtasks. This example demonstrates that it is not always advantageous to schedule each subtask to the processor that offers the minimum EFT, especially when considering dynamic topologies and resources.
\subsubsection{Proposed Degree-based Weighted EFT Mechanism}
To develop our alternative methodology to HEFT, we first define \emph{degree set}.
\begin{myDef}
	\textbf{(Degree set).}  Let $\mathcal{D}^{\bm{\mathsf{VC}}}\left(n_{i},p_{m}\right)$ denotes a degree set containing the vehicles that can receive the data transmission of $\max_{n_{j} \in \mathsf{succ}(n_{i})} \left\{c_{n_{i}, n_{j}}\right\}$ after $n_{i}$ is completed on $p_{m}$, where $\mathsf{succ}(n_{i})$ is the immediate successors set of $n_{i}$, which can be given by
	\begin{equation}
		\begin{aligned}
			\mathcal{D}^{\bm{\mathsf{VC}}}\left(n_{i},p_{m}\right) \triangleq \{p_{n}:\exp \left(-\mathsf{TT}^{\mathsf{max}}_{n_{i}, n_{j}}\left(p_{m}, p_{n}\right) \mu_{p_{m}, p_{n}} \right) \geq \theta, \\ p_{m} \in \mathcal{C}^{\bm{\mathsf{VC}}}(n_{i}),\ p_{n} \in \mathcal{V}^{\bm{\mathsf{VC}}}(n_{j})\}.\label {eq_20}
	\end{aligned}
	\end{equation}
 	And according to (\ref{eq_3}), $\mathsf{TT}^{\mathsf{max}}_{n_{i}, n_{j}}\left(p_{m}, p_{n}\right)$ is the data transmission time of $\max_{n_{j} \in \mathsf{succ}(n_{i})} \left\{c_{n_{i}, n_{j}}\right\}$ between $p_{m}$ and $p_{n}$, expressed by
	\begin{equation}
		\begin{aligned}
		\mathsf{TT}^{\mathsf{max}}_{n_{i}, n_{j}}\left(p_{m}, p_{n}\right) \!\!=\!\! \!\!\!\!\max_{n_{j} \in \mathsf{succ}(n_{i})} \{c_{n_{i}, n_{j}}\} \times{\Gamma\left(\mathsf{PL}\left(d_{p_{m}, p_{n}}\left(\mathsf{st}_{n_{j}}\right)\right)\right)}, \\
		\xi_{n_{i},p_{m}}=1, \ p_{m} \in \mathcal{C}^{\bm{\mathsf{VC}}}(n_{i}), \ p_{n} \in \mathcal{V}^{\bm{\mathsf{VC}}}(n_{j}),    \label{eq_21}		
		\end{aligned}
	\end{equation}
where $\mathsf{succ}(n_{i})$ is the immediate successor set of subtask $n_{i}$.
\end{myDef} 
\begin{algorithm} [!t]
	\centering
	\caption {\textbf{R}anking and \textbf{F}oresight-\textbf{I}ntegrated  \textbf{D}ynamic (RFID) Scheduling Scheme}\label{algorithm_1}
	\begin{algorithmic}[1]
		\STATE \textbf{Input:} $\mathcal{G}^{\bm{\mathsf{A}}}$, $\mathcal{G}^{\bm{\mathsf{VC}}}(n_{\mathsf{entry}})$,  $w_{n_{i}}$, $c_{n_{i},n_{j}}$, $f_{p_{m}}$
		\STATE \textbf{Output:} Scheduling decisions $\{\xi_{n_{i},p_{m}}\}$
		\STATE Schedule subtask $n_{\mathsf{entry}}$ on the task owner
		\WHILE{there are subtask $n_{i} \in \bm{n_{\mathsf{ready}}}$}
		\STATE $\mathsf{s t}_{n_{i}}=\max _{n_{j} \in \mathsf{pred}\left(n_{i}\right)}\left\{\mathsf{AFT}\left(n_{j}\right)\right\}$
		\STATE Calculate the value of $\mathsf{RSRank}^{\mathsf{D}}(n_{i})$ according to (\ref{eq_15})-(\ref{eq_19})
		\STATE Rank $n_{i}\in \bm{n_{\mathsf{ready}}}$ using the value of $\mathsf{RSRank}^{\mathsf{D}}$ in a non-decreasing order
		\STATE $N \leftarrow$ unscheduled subtask with lowest $\mathsf{RSRank}^{\mathsf{D}}$
		\STATE $L \leftarrow$ immediate successor subtasks of $N$
		\FOR {vehicles $p_{m}$ existing in $\mathcal{C}^{\bm{\mathsf{VC}}}\left(N\right)$}
		\STATE Calculate $\mathsf{EFT}(N,p_{m})$ using (\ref{eq_1})-(\ref{eq_9})
		\STATE Calculate $\mathcal{D}^{\bm{\mathsf{CV}}}\left(N,p_{m}\right)$ using (\ref{eq_20})-(\ref{eq_21})
		\STATE Calculate $\mathsf{EFT}^{\mathsf{W}}\left(N, p_{m}\right)$ according to (\ref{eq_22})
		\STATE Return to the beginning of this loop
		\ENDFOR	
		\STATE Schedule subtask $N$ on vehicle $p_{m}$ such that $\mathsf{EFT}^{\mathsf{W}}(N,p_{m})<\mathsf{EFT}^{\mathsf{W}}(N,p_{n}), \ p_{n} \in \mathcal{C}^{\bm{\mathsf{CV}}}\left(N\right) \backslash p_{m}$
		\IF {there are new \emph{ready subtask}s}
		\STATE Return to \textbf{Line 4} 
		\ELSE
		\STATE Go to \textbf{Line 8} 
		\ENDIF
		\ENDWHILE
	\end{algorithmic}
\end{algorithm}
We then assign subtask to the vehicle $p_{m}$ to minimize the weighted EFT:
\begin{equation}
	\begin{aligned}
		\mathsf{EFT}^{\mathsf{W}}\left(n_{i}, p_{m}\right)\!=\!\alpha^{\mathsf{T}} \mathsf{EFT}\left(n_{i}, p_{m}\right)\! - \!\alpha^{\mathsf{R}} \Phi\left(\left|\mathcal{D}^{\bm{\mathsf
			VC}}\left(n_{i},p_{m}\right)\right|\right),
		\\p_{m} \in \mathcal{C}^{\bm{\mathsf{VC}}}\left(n_{i}\right), \label {eq_22}
	\end{aligned}
\end{equation} 
where $\alpha^{\mathsf{T}}$, and $\alpha^{\mathsf{R}}$ represent the preference on completion time and execution success rate (i.e., reliability), respectively. Also, $\Phi()$, which is chosen to be $\Phi\left(\left|\mathcal{D}^{\bm{\mathsf
		VC}}\left(n_{i},p_{m}\right)\right|\right)=0.5\times\left|\mathcal{D}^{\bm{\mathsf
		VC}}\left(n_{i},p_{m}\right)\right|$, is a monotone increasing function. Specifically, in (\ref{eq_22}) we not only consider the performance (in terms of completion time) when $n_{i}$ is assigned to $p_{m}$, but also look ahead to the number of vehicles (i.e., $\left|\mathcal{D}^{\bm{\mathsf{VC}}}\left(n_{i},p_{m}\right)\right|$) the can receive the data transmission of $\max_{n_{j} \in \mathsf{succ}(n_{i})} \left\{c_{n_{i}, n_{j}}\right\}$, after $n_{i}$ is completed on $p_{m}$. Note that a larger value of $\left|\mathcal{D}^{\bm{\mathsf{VC}}}\left(n_{i},p_{m}\right)\right|$ can lead to a lower value for $\mathsf{EFT}^{\mathsf{W}}\left(n_{i}, p_{m}\right)$ (i.e.,  a vehicle with general computation ability but commendable transmission potential can be given with high weight).

\subsection{Summary}
In RFID, the scheduling priority of each subtask relies heavily on the selected vehicles of its immediate predecessor subtasks, which is also influenced by the current resources supply. Then, a degree-based weighted EFT is leveraged to assign subtask to the vehicle, which can offer faster processing with reliable transmission links.

\textbf{Algorithm 1} shows the details of RFID. It can be verified that the time complexity of RFID is $\bm{\mathcal{O}}((\text{n}+\text{r})^2\cdot \text{p}^2)$ where $n$, $r$, and $p$ are the number of subtasks within the DAG task, the maximum number of successors per DAG task, and the maximum number of vehicles involved in the VC, respectively. 
\subsection{A Toy Example}
\begin{figure}[H]
	\centering
	\includegraphics[width=3.5in]{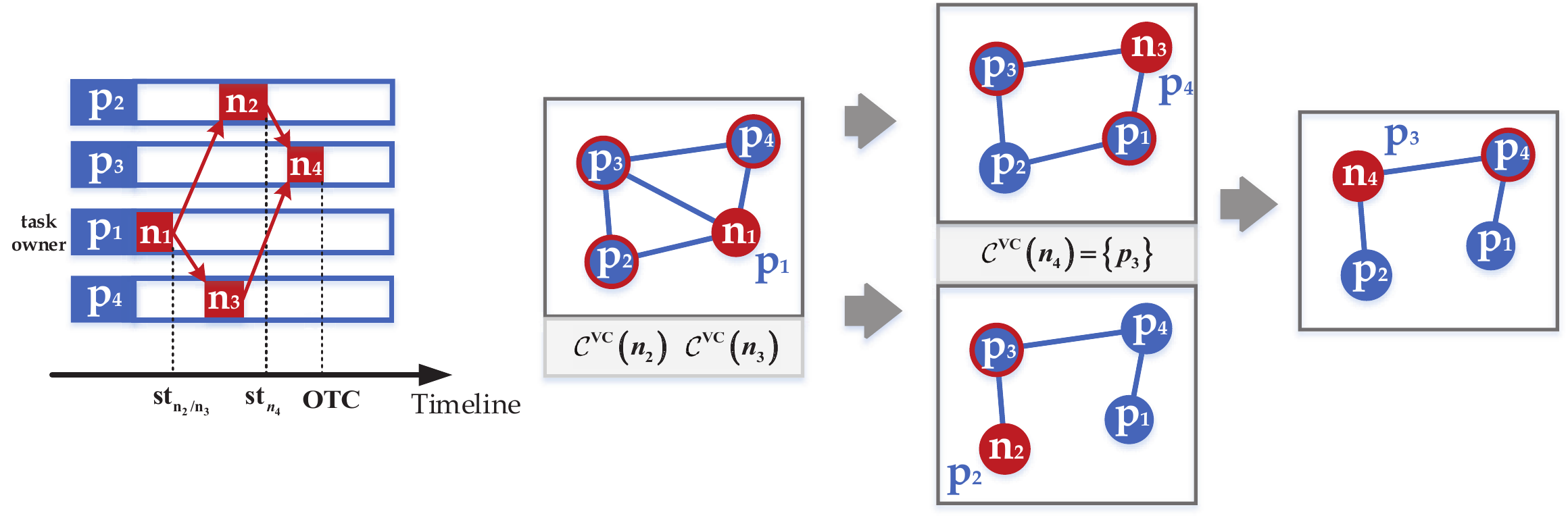}
	\caption{Subtask-vehicle mapping.}
	\label{fig_7}
\end{figure}
A feasible mapping among subtasks and vehicles associated with VC$_{2}$ (given by Fig. \ref{fig_2}) is shown in Fig. \ref{fig_7}, where the left subplot represents a specific processing procedure (i.e., the assignment of subtasks to vehicles); while the right subplot depicts the changing topology of VC$_{2}$ and corresponding $\mathcal{C}^{\bm{\mathsf{VC}}}(n_{i})$ during the scheduling procedure. Since $n_{1}$ (i.e., $n_{\mathsf{entry}}$) is scheduled on $p_{1}$ (i.e., task owner), subtasks $n_{2}$ and $n_{3}$ become \emph{ready subtask}s. After sorting $n_{2}$, $n_{3}$ under resource-scarcity based dynamic downward ranking mechanism, $n_{3}$ is scheduled firstly to its most desirable vehicle, based on its candidate vehicle set $\mathcal{C}^{\bm{\mathsf{VC}}}(n_{2})=\{p_{1}, p_{2}, p_{3}, p_{4}\}$. Then, according to the degree-based weighted EFT, $n_{3}$ and $n_{2}$ are scheduled on $p_{4}$ and $p_{2}$, respectively. Finally, subtask $n_{4}$ becomes \emph{ready subtask}, and is scheduled on vehicle $p_{3}$, which can receive the output data both of $n_{2}$ and $n_{3}$.

\section{Performance Evaluation}
We conduct comprehensive simulations to evaluate the performance of RFID. To quantify the performance of our proposed methodology, we consider three key performance metrics: \emph{\romannumeral1}) overall DAG task completion time, \emph{\romannumeral2}) execution success rate, and \emph{\romannumeral3}) running time of the algorithm.
\subsection{Simulation Setup}
\begin{figure}[!t]
	\centering
	\subfigure[]{
		\begin{minipage}[t]{0.5\linewidth}
			\centering
			\includegraphics[width=\linewidth]{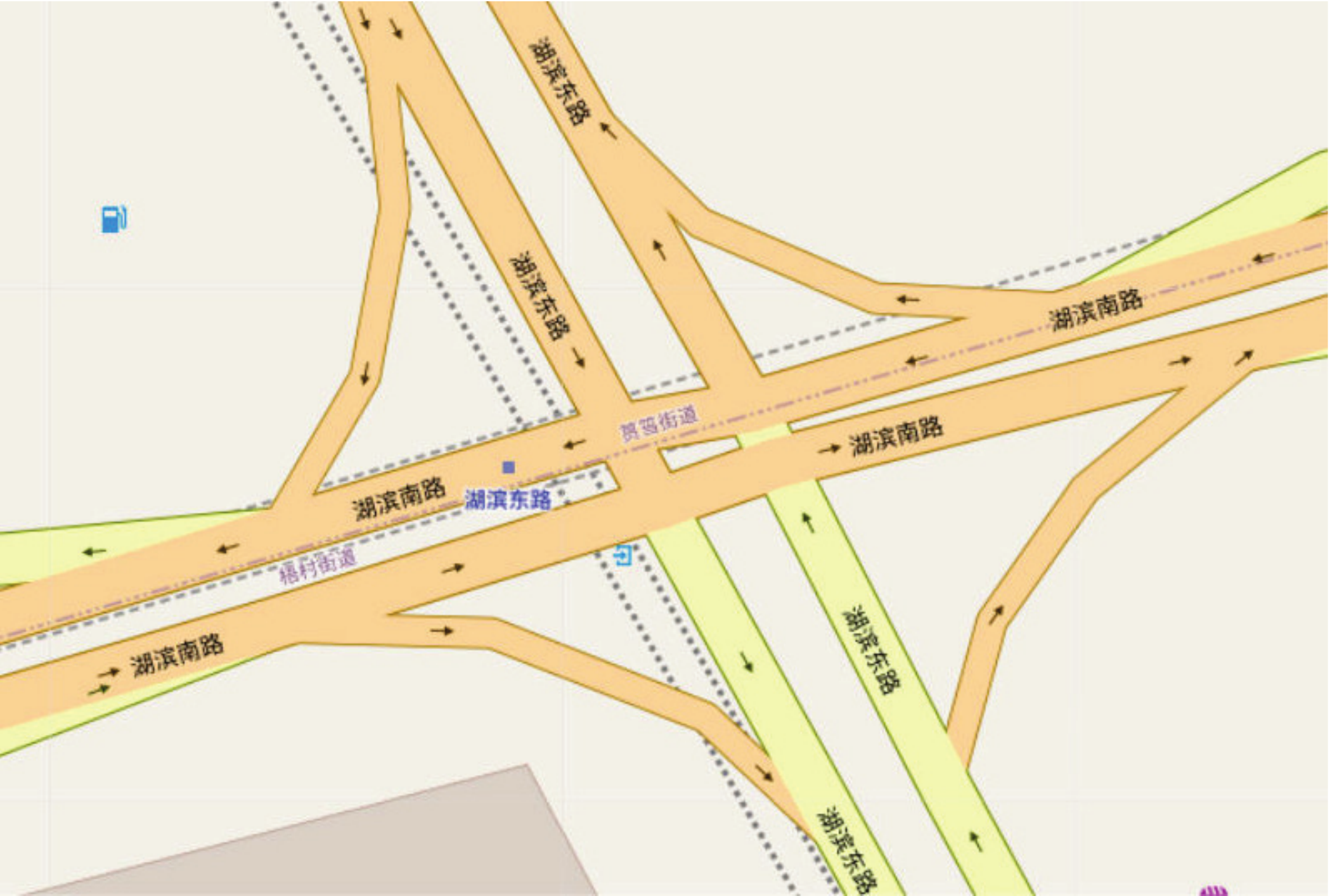}
			%\caption{fig1}
		\end{minipage}%
	}%
	\subfigure[]{
		\begin{minipage}[t]{0.5\linewidth}
			\centering
			\includegraphics[width=0.85\linewidth]{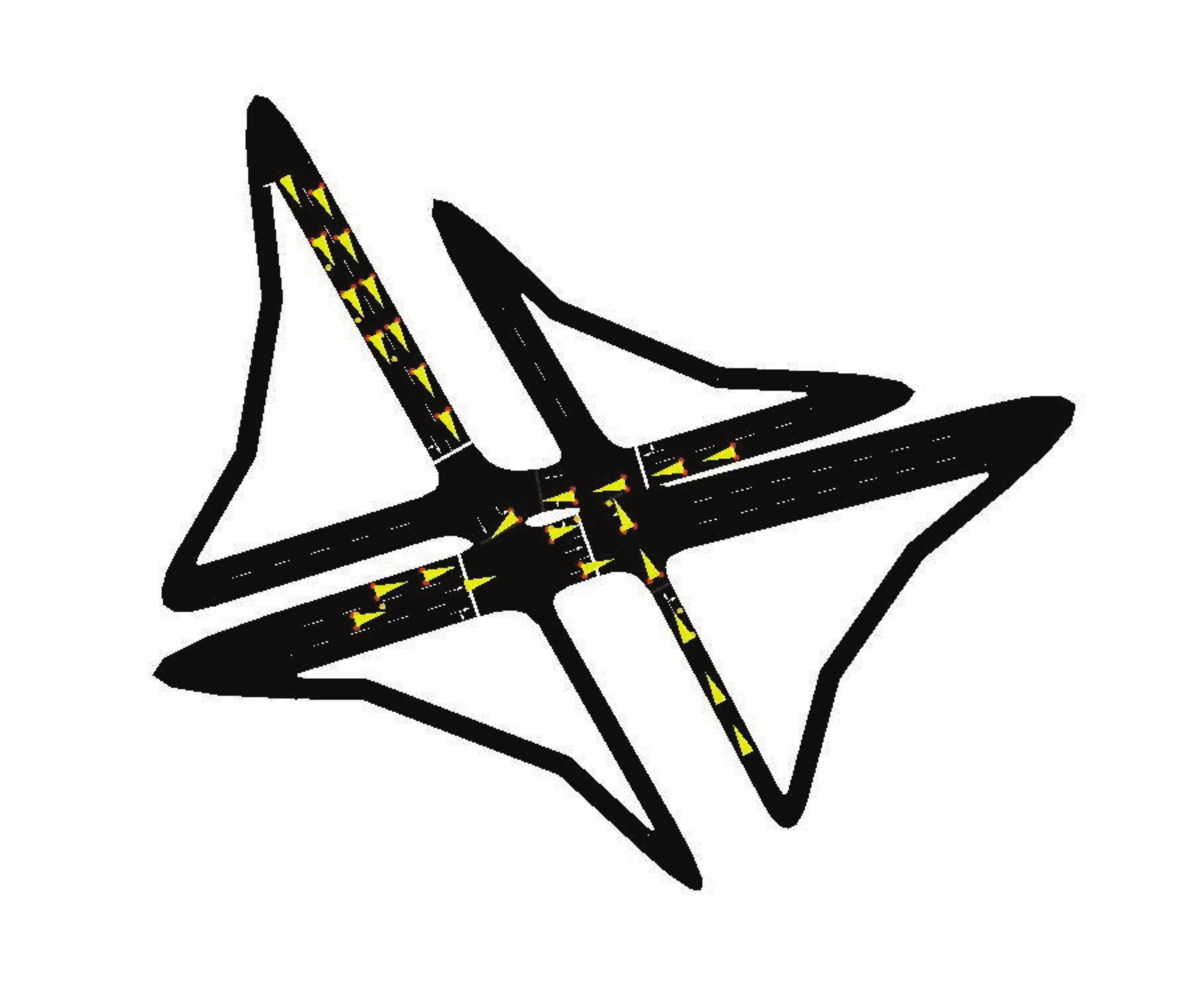}
			%\caption{fig2}
		\end{minipage}%
	}%
	\centering
	\caption{VC network visualization.}
	\label{fig_8}
\end{figure}

\emph{Parameter setting of VC:} We consider a real-world traffic region (shown in Fig. \ref{fig_8}(a)) with size of 1\textrm{km} × 1\textrm{km} in Xiamen, Fujian, China, obtained from OpenStreetMap\cite{b55}. SUMO\cite{b56} is utilized to import mobile vehicles and subsequently form a realistic VC over the simulation region in Fig. \ref{fig_8}(b). Assuming that vehicles located within 500\textrm{m} from each other are connected via one-hop V2V links and form an undirected graph (VC). We conduct simulations upon considering different number of vehicles to better capture various vehicle density associated with a VC, and different VC's topologies. Each vertex in the VC graph represents a vehicle with certain computing capability, which follows a normal distribution with mean 20\textrm{MHz} and variance 0.2\cite{b54}. The weight of each link (edge) connecting two vertices represents the residual contact duration. In addition, a monotone increasing function $\Gamma(\mathsf{PL}\left(d_{p_{m}, p_{n}}\left(t\right)\right)) = 0.15 \times \mathsf{PL}\left(d_{p_{m}, p_{n}}\left(t\right)\right) + 0.001$ is applied to determine the transmission time between different vehicles.

\emph{Parameter setting of DAG task:} We implement a DAG generator \cite{b22} to randomly generate different types of DAG task, regarding the number of subtasks, communication-to-computation ratio (CCR) of subtasks, and the number of layers. Specifically, subtask belongs to the same layer can be processed in parallel and each layer in the DAG contains at least one subtask. Besides, the first and the last layer are occupied only by $n_{\mathsf{entry}}$, and $n_{\mathsf{exit}}$, respectively. The computing workload of each subtask obeys a normal distribution with mean 3\textrm{Mclock cycles} and 0.2 variance\cite{b21}. The corresponding transmission data size of each edge follows a normal distribution with mean 1.2\textrm{Mbit} and variance 0.2\cite{b21}. Also, we consider a real DAG task, i.e., molecular dynamics code DAG, to achieve better performance evaluation, which has been adopted in many existing works, such as \cite{b27}, \cite{b40}.
\subsection{Benchmarks}
To better verify the performance of our proposed RFID, three benchmarks are considered below, as inspired by some existing works:
\begin{itemize}
	\item {\emph{Heterogeneous Earliest Finish Time (HEFT) \cite{b40}:} HEFT firstly ranks all the subtasks according to their average completion time, which is computed recursively from the entry subtask, i.e., conventional \emph{downward ranking}. Then, the subtask with the highest scheduling priority is assigned to the vehicle that can process the subtask in the shortest time without considering the corresponding transmission constraints (\ref{eq_14}).}
\end{itemize}
\begin{itemize}
	\item {\emph{Lookahead (LA)\cite{b15}:} As an improvement of HEFT, LA ranks subtasks similar to HEFT. Then, it assigns subtask $n_{i}$ with the highest scheduling priority to vehicle $p_{m}$, which can minimize the maximum completion time of subtask $n_{j}$, where $n_{j} \in \mathsf{succ}(n_{i})$, after $n_{i}$ completes on $p_{m}$. In this paper, one-step LA is considered; namely, we only care about the immediate successors of each current scheduling subtask.}
\end{itemize}
\begin{itemize}
	\item {\emph{Modified Genetic Algorithm (MGA) \cite{b24}:} MGA consists of three key parts. First, an integer encoding is employed to denote the assignment between subtasks and vehicles. Then, to guarantee the feasibility of generated solutions, the definition of relatives (that is, two subtasks with dependency) is used to avoid the mis-operation of crossover (a subtask cannot be scheduled earlier than its predecessor). Finally, the mutation is adopted to improve the fitness of the candidate solutions.}
\end{itemize} 

\begin{figure*}[!t]
	\centering
	\subfigure[]{
		\begin{minipage}[t]{0.45\linewidth}
			\centering
			\includegraphics[width=\linewidth]{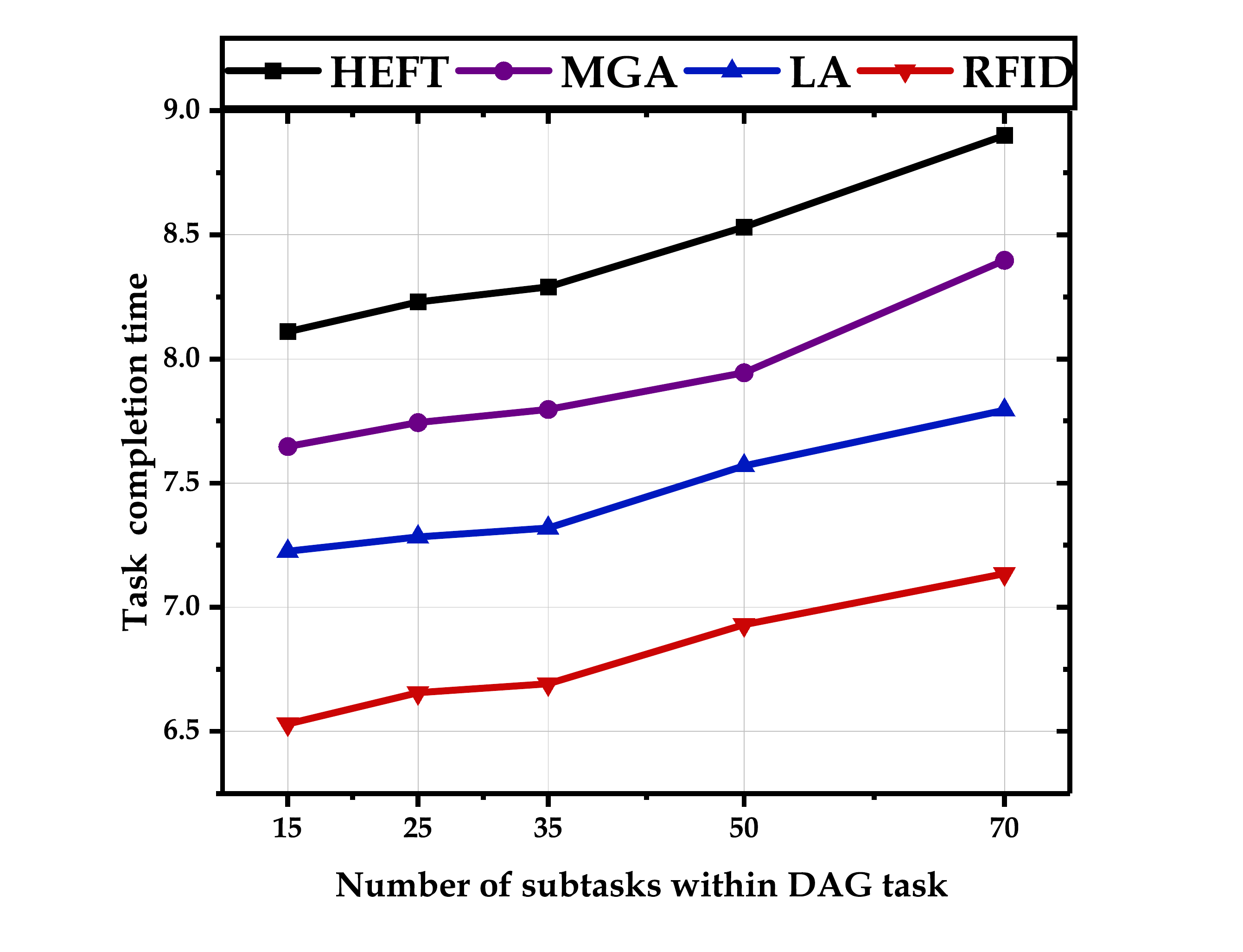}
			%\caption{fig1}
		\end{minipage}%
	}%
	\subfigure[]{
		\begin{minipage}[t]{0.45\linewidth}
			\centering
			\includegraphics[width=\linewidth]{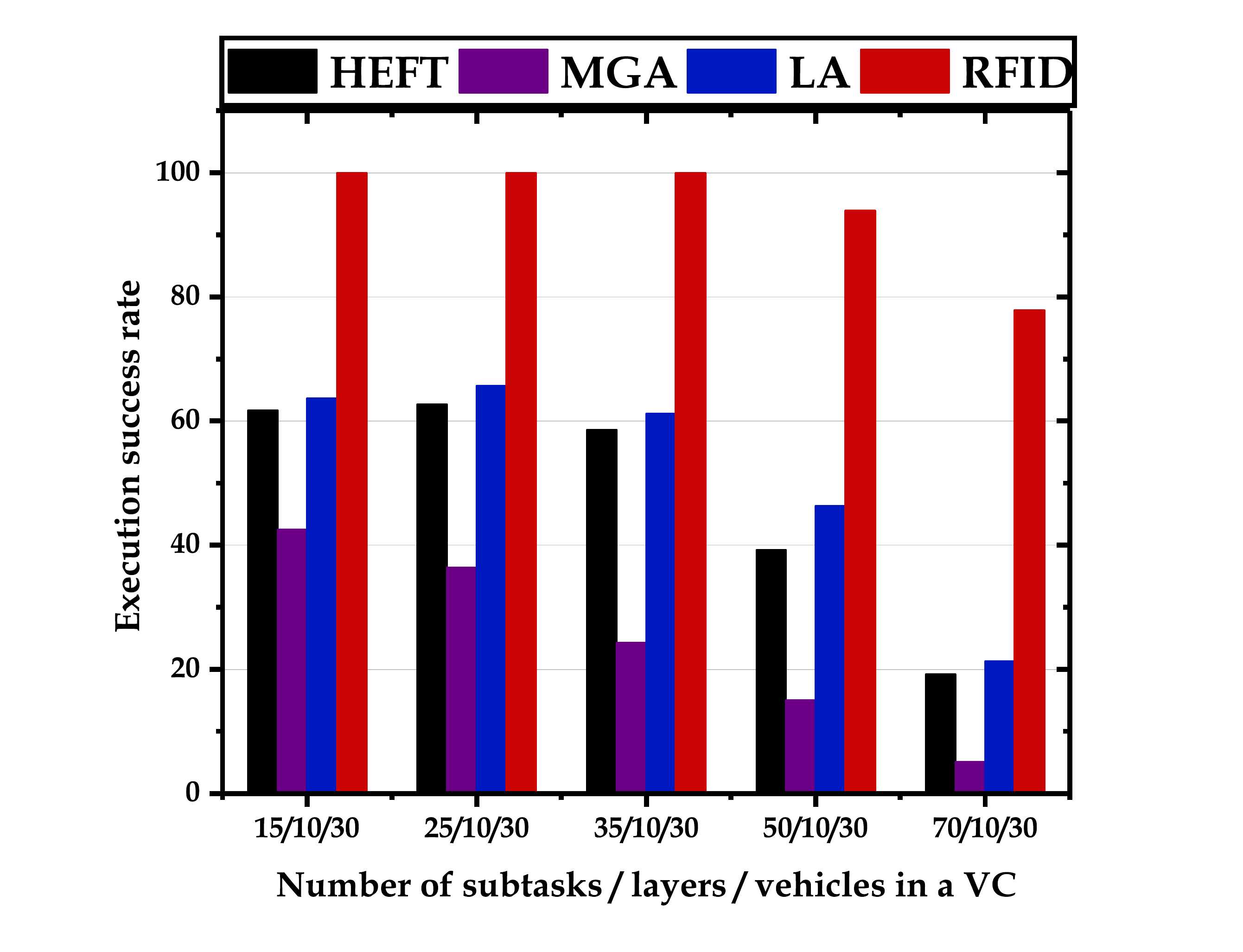}
			%\caption{fig2}
		\end{minipage}%
	}%
	\centering
	\caption{Performance evaluations upon considering different number of subtasks within DAG task, for randomly generated DAG tasks. }
	\label{fig_9}
\end{figure*} 
\begin{figure*}[!t]
	\centering
	\subfigure[]{
		\begin{minipage}[t]{0.45\linewidth}
			\centering
			\includegraphics[width=\linewidth]{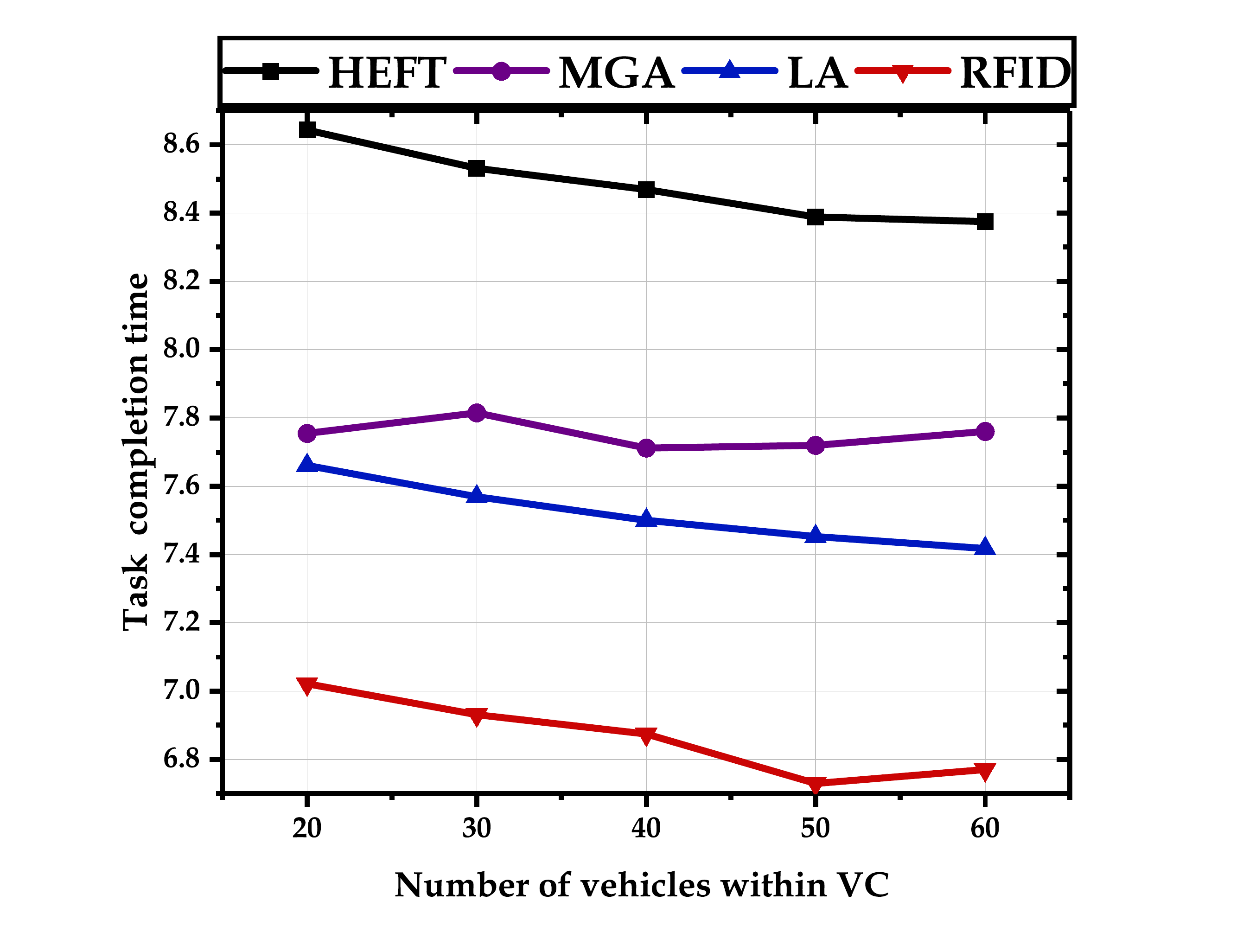}
			%\caption{fig1}
		\end{minipage}%
	}%
	\subfigure[]{
		\begin{minipage}[t]{0.45\linewidth}
			\centering
			\includegraphics[width=\linewidth]{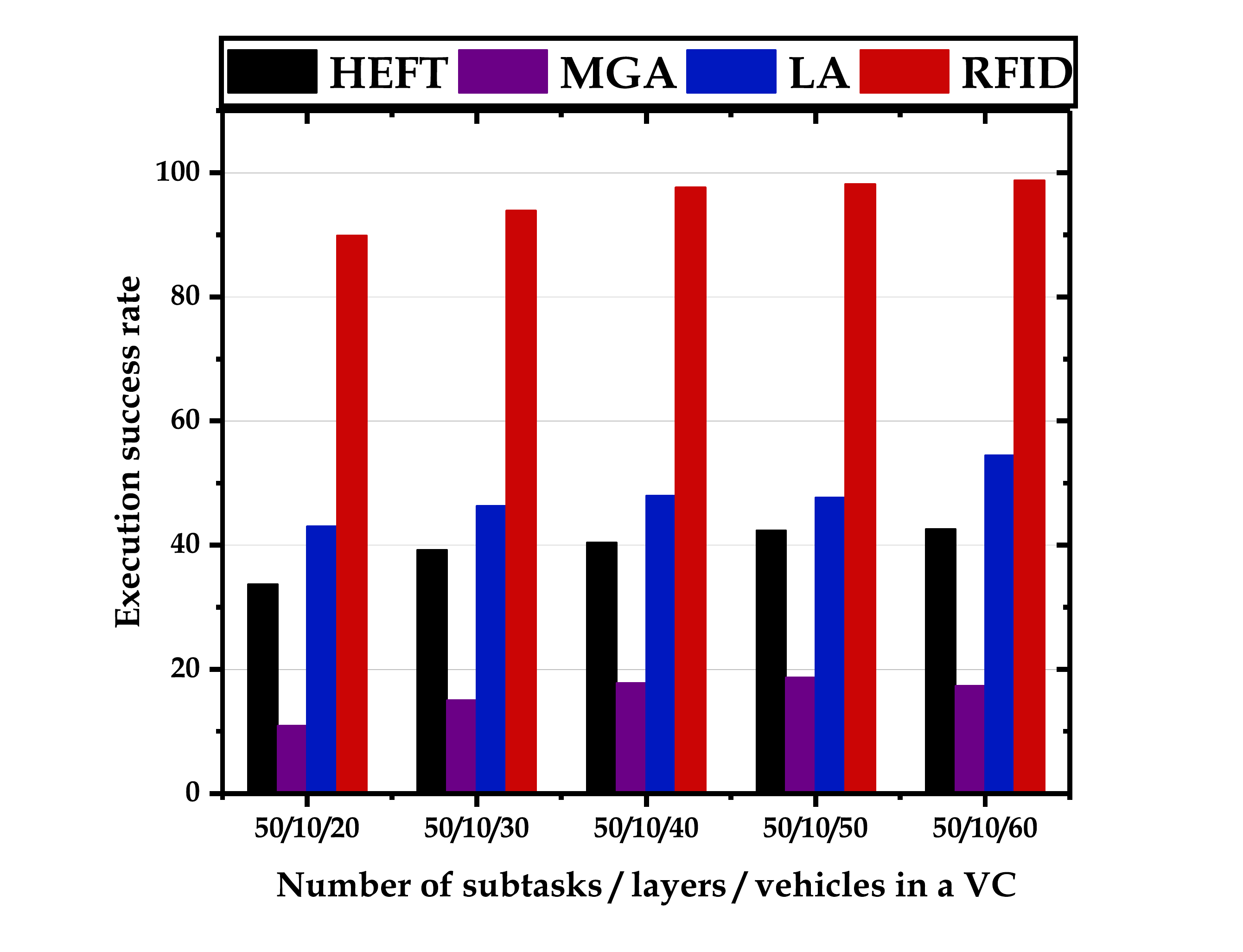}
			%\caption{fig2}
		\end{minipage}%
	}%
	\centering
	\caption{Performance evaluations upon considering different number of vehicles within VC, for randomly generated DAG tasks.}
	\label{fig_10}
\end{figure*} 
\subsection{Simulation Results of Randomly Generated DAG Tasks}
Performance comparisons and evaluations are conducted, in terms of average completion time and execution success rate, by considering diverse numbers of subtasks, vehicles, layers of DAG tasks, and CCR of the subtask. Besides, 1000 independent iterations are simulated, as benefited by the Monte Carlo method.

\emph{1) Impacts of diverse numbers of subtasks:} Fig. \ref{fig_9}(a) shows the performance evaluation of the overall DAG task completion time, with an increasing number of subtasks (from 15 to 70 ). The number of layers is set as 10; CCR is set by 1, and the number of vehicles involved in the initial VC is set as 30. As can be seen from Fig. \ref{fig_9}(a), our proposed RFID outperforms the other three baseline methods with a much faster task completion time. Specifically, since low resource demands can lead to weak resource competition among vehicles, the curve of task completion time between 25 subtasks and 35 subtasks is relatively flat. Compared to LA, although completion time is not the only indicator for optimization, RFID can reduce the overall DAG task completion time with the help of changing the scheduling priority of some specific subtasks according to their resource availability. In summary, the performance improvement of RFID in terms of task completion time is 19.48$\%$ better than HEFT, 14.61$\%$ better than MGA, 9.63$\%$ better than LA at 15 subtasks; and is 19.83$\%$ better than HEFT, 15.02$\%$ better than MGA, and 8.43$\%$ better than LA at 70 subtasks.

\begin{figure*}[!t]
	\centering
	\subfigure[]{
		\begin{minipage}[t]{0.45\linewidth}
			\centering
			\includegraphics[width=\linewidth]{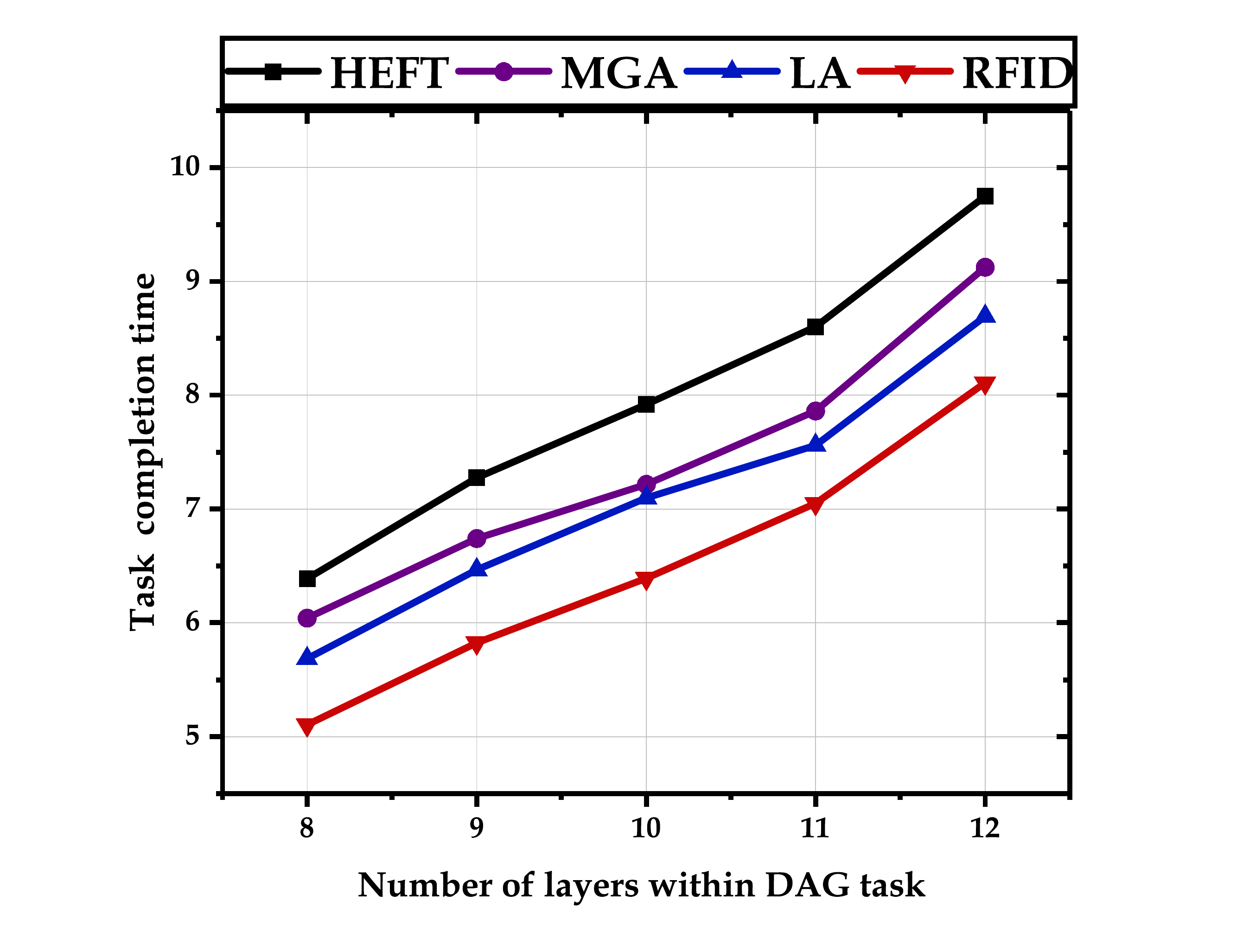}
			%\caption{fig1}
		\end{minipage}%
	}%
	\subfigure[]{
		\begin{minipage}[t]{0.45\linewidth}
			\centering
			\includegraphics[width=\linewidth]{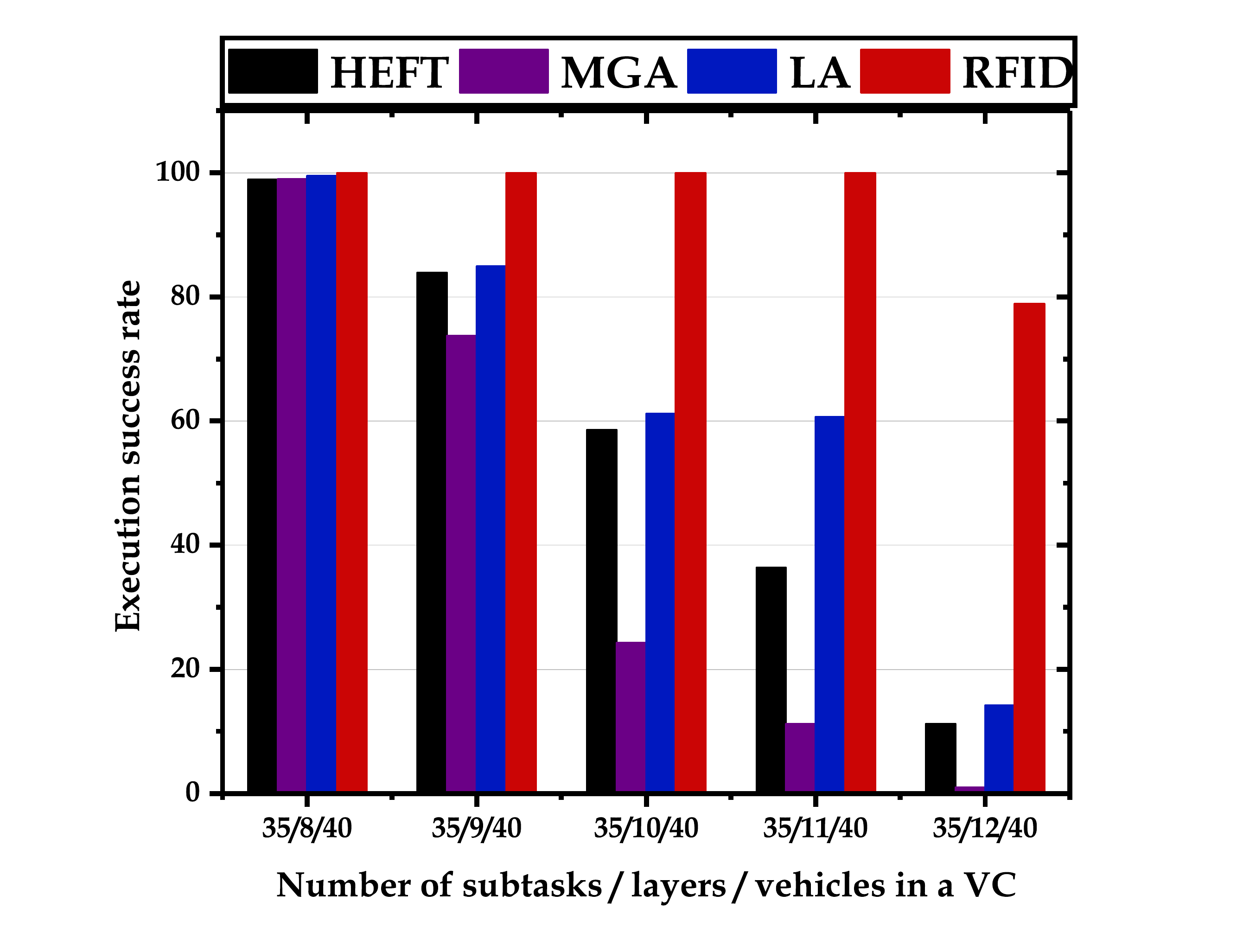}
			%\caption{fig2}
		\end{minipage}%
	}%
	\centering
	\caption{Performance evaluations upon considering different number of layers, for randomly generated DAG tasks.}
	\label{fig_11}
\end{figure*} 
\begin{figure*}[!t]
	\centering
	\subfigure[]{
		\begin{minipage}[t]{0.45\linewidth}
			\centering
			\includegraphics[width=\linewidth]{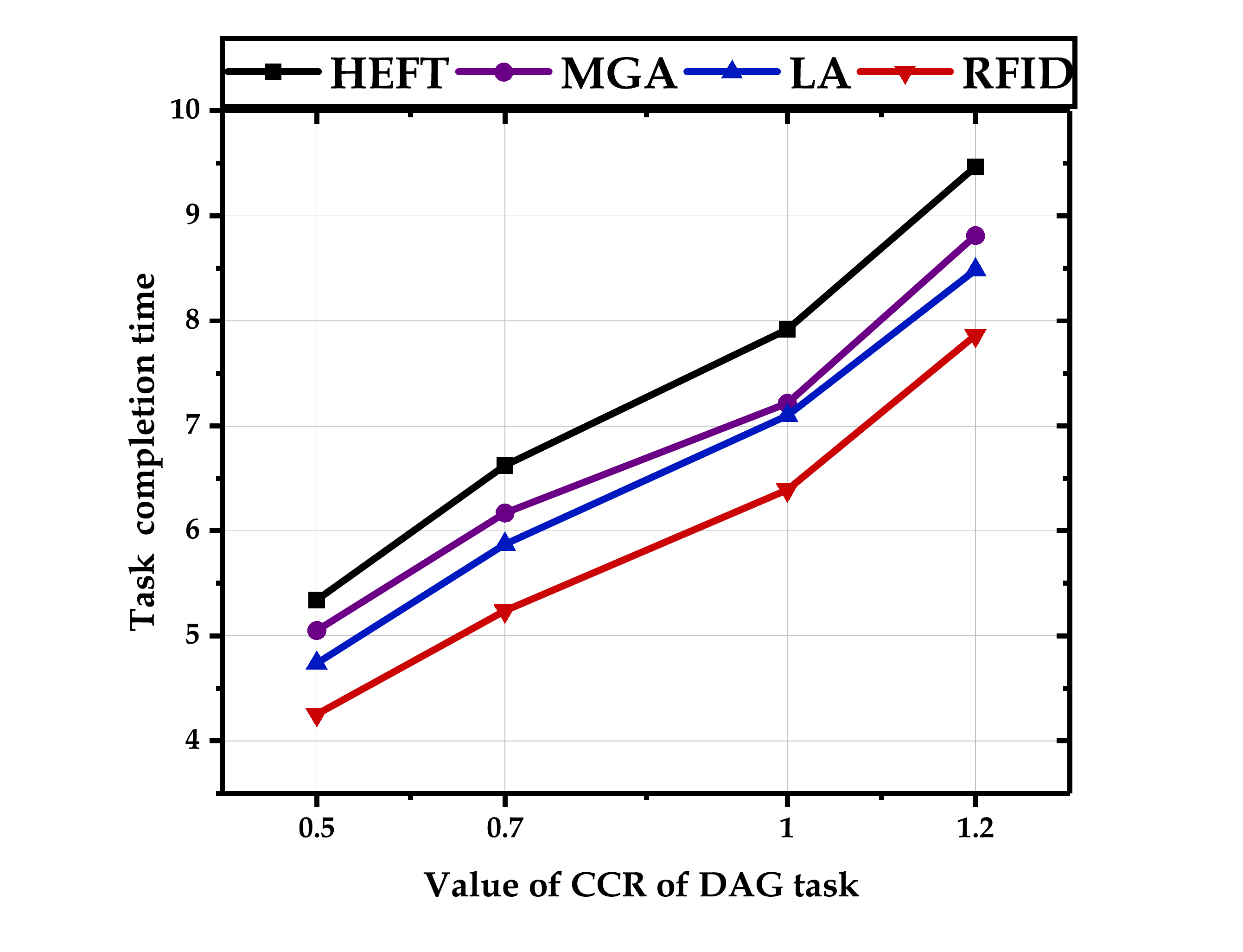}
			%\caption{fig1}
		\end{minipage}%
	}%
	\subfigure[]{
		\begin{minipage}[t]{0.45\linewidth}
			\centering
			\includegraphics[width=\linewidth]{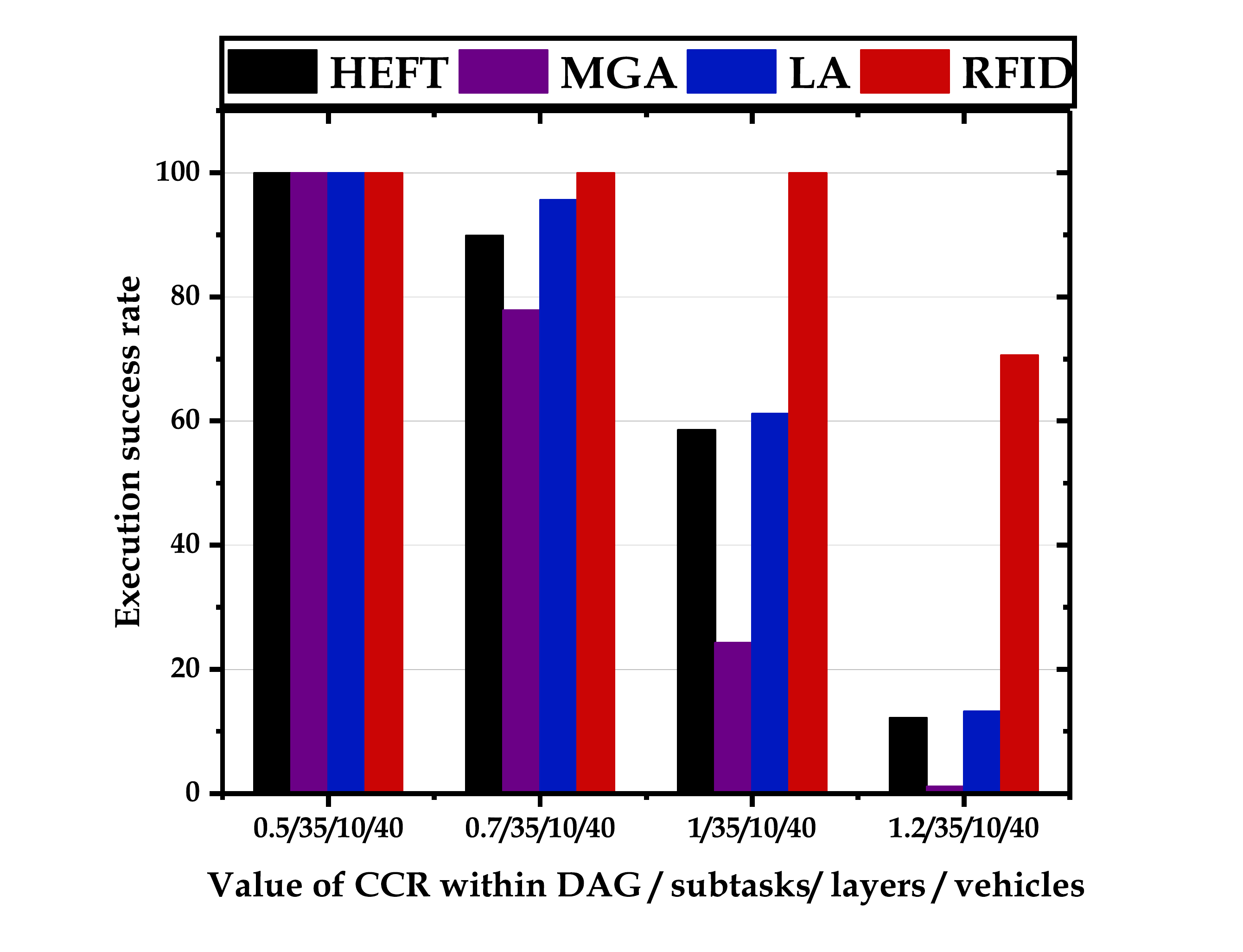}
			%\caption{fig2}
		\end{minipage}%
	}%
	\centering
	\caption{Performance evaluations upon considering various CCR, for randomly generated DAG tasks.}
	\label{fig_12}
\end{figure*}

Fig. \ref{fig_9}(b) illustrates the performance with an increasing number of subtasks (from 15 to 70) on the execution success rate. It can be observed that RFID significantly increases the corresponding execution success rate. Notably, there exists a decreasing execution success rate in the curve of HEFT, MGA, and LA, mainly due to the increasing number of subtasks, which requires heavily V2V connection time (i.e., the higher requirement on residual V2V contact duration). Besides, in Fig. \ref{fig_9}(b), MGA has the worst performance on execution success rate because of its randomly generated solution space. In comparison with HEFT, LA slightly enhances the execution success rate by considering the impact of scheduling the successors of each assigned subtask. Also, RFID outperforms LA by increasing the probability of scheduling each subtask to the reliable connected vehicles by considering the degree of each vehicle. In summary, the performance improvement of RFID in terms of success rate is 38.3$\%$ better than HEFT, 57.5$\%$ better than MGA, 36.6$\%$ better than LA at 15 subtasks; and is 58.7$\%$ better than HEFT, 72.8$\%$ better than MGA, and 56.6$\%$ better than LA at 70 subtasks.

\emph{2) Impacts of diverse numbers of vehicles involved in VC:}
Fig. \ref{fig_10}(a) demonstrates the performance by increasing the number of vehicles from 20 to 60 on the overall DAG task completion time. The number of layers is set as 10; the CCR is set by 1; and the number of subtask is set by 50. It can be seen from Fig. \ref{fig_10}(a) that the increasing number of vehicles can significantly accelerate the completion of the DAG tasks thanks to more sufficient resources. However, after more than 50 vehicles, task completion time can not be reduced significantly, even slightly increasing. This phenomenon mainly owes to the redundancy of available resources complicating VC's topology, which thus may cause subtasks to be assigned to vehicles with poor computing capability. Specifically, the growing resource supply impacts less on MGA, since each subtask is randomly scheduled on the vehicles at the beginning, while the chromosome crossing procedure of MGA may fail to ensure the feasibility of the newly generated solution (i.e., the connection constraints of newly assigned subtasks are not taken into account in advance). To this end, an increasing number of vehicles in MGA does not imply an increase in the number of feasible solutions, which thus fails to bring an obvious decrease in task completion time. In summary, the performance improvement of RFID in terms of the task completion time is 18.76$\%$ better than HEFT, 9.45$\%$ better than MGA, 8.33$\%$ better than LA at 20 vehicles; and is 19.16$\%$ better than HEFT, 12.76$\%$ better than MGA, and 8.72$\%$ better than LA at 60 vehicles. 

Fig. \ref{fig_10}(b) evaluates the performance of execution success rate upon considering the different numbers of vehicles involved in a VC. It can be observed that an increasing number of vehicles can bring a larger execution success rate due to more available resources. The performance trends of MGA, HEFT, and LA in terms of the execution success rate are similar to that of Fig. \ref{fig_10}(b). Specifically, the performance improvement of RFID in terms of the execution success rate is 56.2$\%$ better than HEFT, 79$\%$ better than MGA, 46.9$\%$ better than LA at 20 vehicles; and is 56.2$\%$ better than HEFT, 81.5$\%$ better than MGA, and 44.3$\%$ better than LA at 60 vehicles.  

\emph{3) Impacts of diverse numbers of layers of DAG task:}  
Fig. \ref{fig_11}(a) shows the impacts of changing the number of layers from 8 to 12 on the overall DAG task completion time. The number of subtasks is set by 35; CCR is set as 1, and the number of vehicles is set by 40. Interestingly, increasing the number of layers can significantly lead to a larger overall DAG task completion time due to unsatisfying parallelism of the DAG task. Specifically, when the parallelism of a DAG task keeps decreasing (namely, the number of layers keeps increasing), more subtasks require sequential execution, which causes an increase in the completion time of the DAG task. Notably, when the number of layers of a DAG task equals that of subtasks (e.g., chess game task shown in Fig. \ref{fig_1}), local computing will become the best choice to avoid frequent data transmission among vehicles. In addition, as the number of layers increases, MGA and LA perform almost equivalently, which indicates that the decrease of the average number of successors of each subtask can result in an ineffectiveness when adopting LA, which concerns the impact of scheduling of the successors of the each assigned subtask. Specifically, the performance improvement of RFID in terms of the overall DAG task completion time is 20.08$\%$ better than HEFT, 15.5$\%$ better than MGA, 10.22$\%$ better than LA at 8 layers, and is 16.82$\%$ better than HEFT, 11.15$\%$ better than MGA, and 6.74$\%$ better than LA at 12 layers. 

Fig. \ref{fig_11}(b) depicts the execution success rate performance upon considering different layers (from 8 to 12) in a DAG task. It can be observed from Fig. \ref{fig_11}(b) that all the four algorithms show a decreasing trend in execution success rate as the number of layers increases, especially for HEFT, MGA, and LA. The key reasons are that increasing the number of layers can lead to increased data transmission time between interdependent subtasks and ready time at the vehicles. For example, given a small number of layers (e.g., one layer contains many subtasks), most subtasks can be processed in parallel, which thus results in fast completion. More importantly, the VC's topology during task scheduling may stay relatively stable due to the high parallelism of the DAG task. Using MGA can cause sharp performance degradation as the number of layers changes from 9 to 10, which indicates that layers heavily impact the random-based subtask-vehicle assignment mechanism. In summary, the performance improvement of RFID in terms of the execution success rate is 1.1$\%$ better than HEFT, 1$\%$ better than MGA, 0.5$\%$ better than LA at 8 layers; and is 67.7$\%$ better than HEFT, 78.8$\%$ better than MGA, and 64.7$\%$ better than LA at 12 layers.

\begin{figure}[!t]
	\centering
	\includegraphics[width=3in]{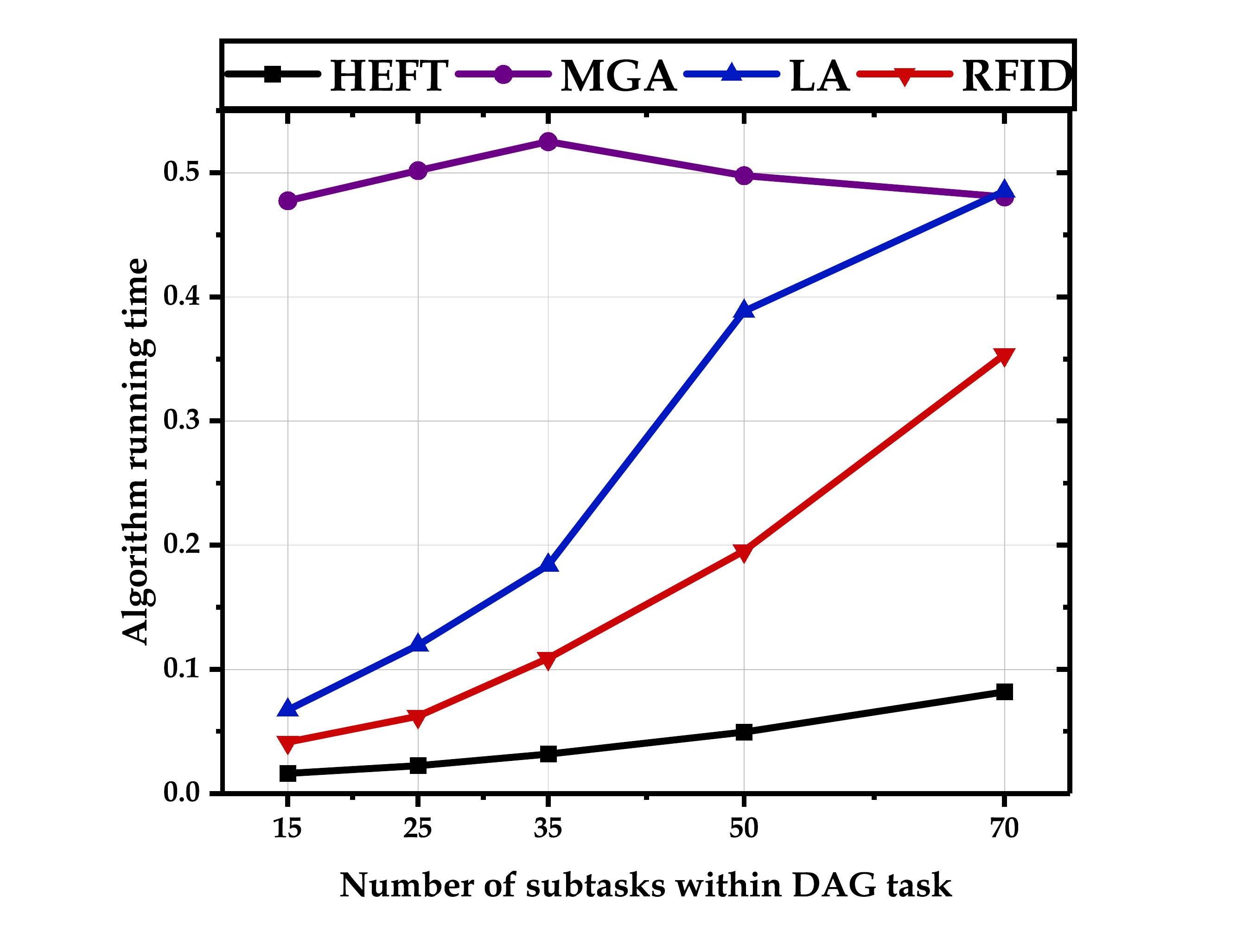}
	\caption{Performance evaluations upon considering different number of subtasks associated with DAG task, on algorithm's running time. }
	\label{fig_13}
\end{figure}
\begin{figure}[!t]
	\centering
	\includegraphics[width=3in]{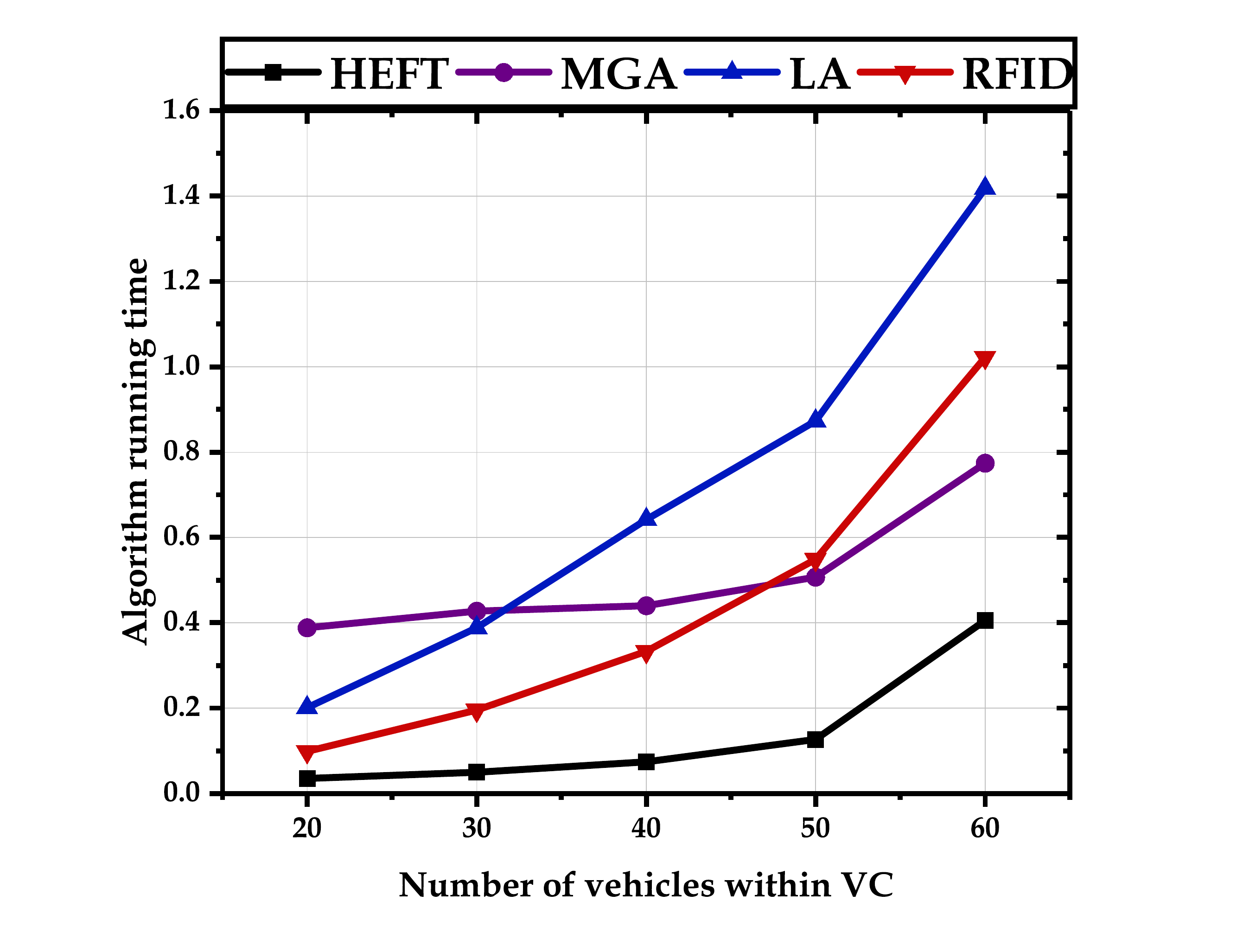}
	\caption{Performance evaluations upon considering different number of vehicles associated with VC, on algorithm's running time.}
	\label{fig_14}
\end{figure}

\emph{4) Impacts of considering various CCR of DAG Task:}
Different values of CCR are tested in Fig. \ref{fig_12}(a). The number of subtasks and layers are set by 35 and 10, respectively, while the number of vehicles is set as 40. As can be seen from Fig. \ref{fig_12}(a), the overall DAG task completion time rises as the CCR of each subtask increases due to the growing data transmission time. Additionally, MGA and LA show similar performance trends in Fig. \ref{fig_11}(a), which indicates that an increase in the CCR of the DAG task conforms to a decrease in the parallelism of the DAG task and thus results in a significant raising in data transmission time. In summary, the performance improvement of RFID in terms of the task completion time is 20.39$\%$ better than HEFT, 15.79$\%$ better than MGA, 10.23$\%$ better than LA at 0.5 CCR; and is 16.95$\%$ better than HEFT, 10.77$\%$ better than MGA, and 7.34$\%$ better than LA at 1.2 CCR.

Fig. \ref{fig_12}(b) compares the performance on execution success rate by considering CCR from 0.5 to 1.2. Similar to Fig. \ref{fig_11}(b), given a small CCR, due to the fast data transmission, VC's topology stays relatively stable during the execution of subtasks, and all the algorithms can enjoy a high execution success rate. In summary, the performance improvement of RFID in terms of the execution success rate is as same as HEFT, MGA, and LA at 0.5 CCR and is 58.4$\%$ better than HEFT, 69.5$\%$ better than MGA, and 57.4$\%$ better than LA at 1.2 CCR.

\subsection{Comparison of Running Time of Different Algorithms} 
In this subsection, we conduct a performance comparison in terms of the algorithm running time by considering various numbers of subtasks associated with randomly generated DAG tasks and different vehicle densities. 

\emph{1) Impacts of diverse numbers of subtasks:}
Fig. \ref{fig_13} shows the impact on algorithm running time caused by a varying numbers of subtasks within a DAG task (from 15 to 70). The CCR of the DAG task is set by 1; the number of layers is set as 10, and the number of the vehicle is considered by 30. Due to the predefined initial population number of MGA (mainly used to generate random assignments), the running time of MGA is slightly affected by increasing the number of subtasks. Besides, in comparison with the other three algorithms, MGA has the longest running time caused by the internal iteration time for convergence. Also, HEFT outperforms other algorithms in running time since increasing the number of subtasks can lead to an extra increasing time in evaluating the performance of corresponding successors, which significantly raises the running time of LA. Since RFID only needs to evaluate the transmission data size between different subtasks and successors, the corresponding running time remains relatively lower as compared to LA. In summary, although the running time of RFID is higher than that of HEFT, its overall DAG task completion time and execution success rate significantly outperforms HEFT, as verified by Fig. \ref{fig_9}-Fig. \ref{fig_12}.

\emph{2) Impacts of diverse number of vehicles:}
Fig. \ref{fig_14} compares that performance upon considering a varying number of vehicles (from 20 to 60) on the algorithm's running time. The number of subtasks and layers are set by 50 and 10, respectively, while the CCR of the DAG task is set as 1. The growing vehicles can bring rising running time since more resource providers should be considered during task scheduling. Specifically, LA's running time is significantly impacted by the number of vehicles, which mainly owes to the increased time spent on evaluating the performance of each vehicle to determine the assignment of each subtask. Similarly, in RFID, more vehicles can complicate the topology of VC, resulting in an increased running time for calculating the degree of each vehicle. In summary, although the running time of RFID is higher than that of HEFT, its overall DAG task completion time and the execution success rate is significantly better than HEFT as verified in Fig. \ref{fig_9}-Fig. \ref{fig_12}.
\begin{figure}[!t]
	\centering
	\includegraphics[width=3.5in]{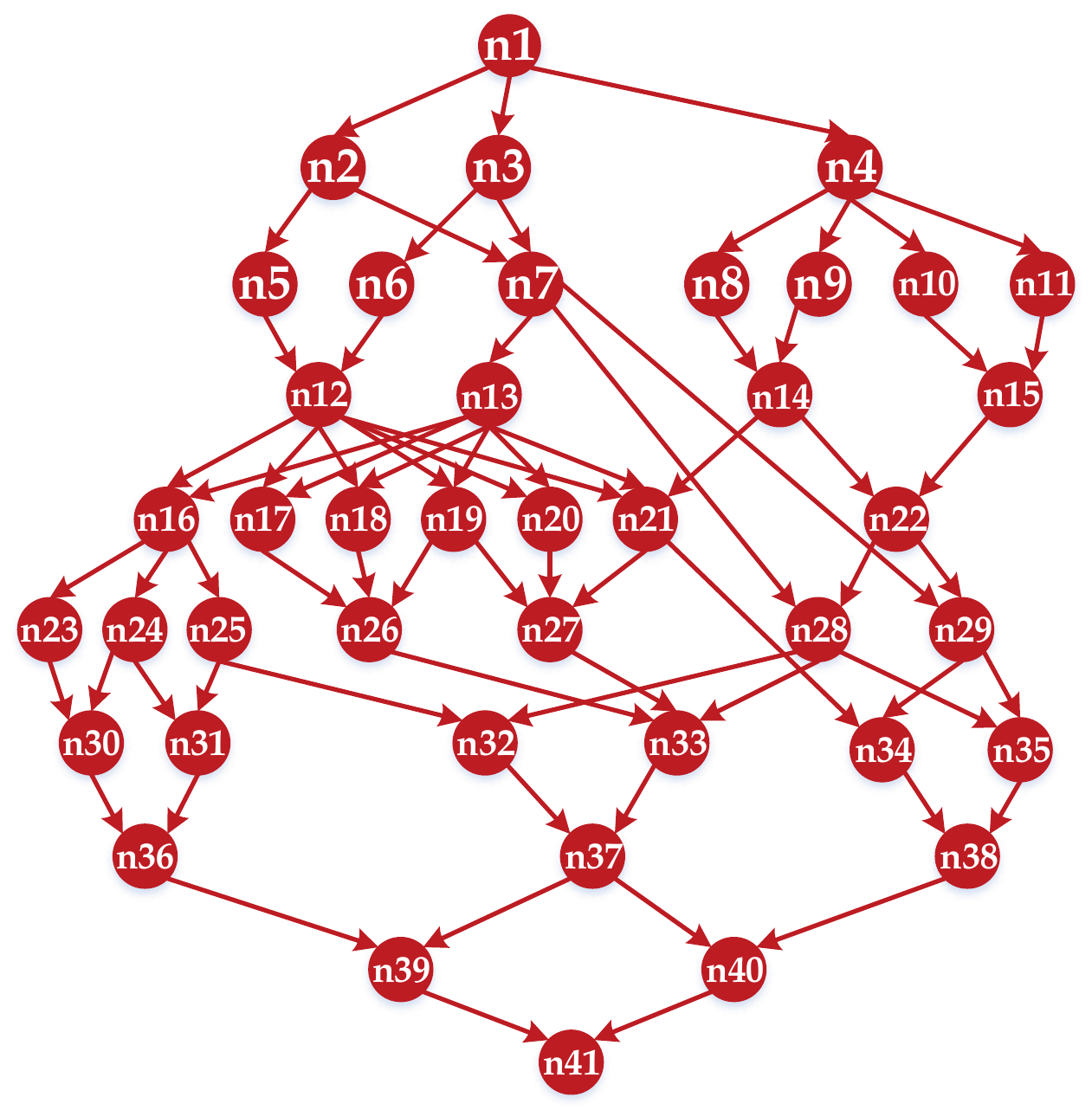}
	\caption{The DAG of the molecular dynamics code. \cite{b27}\cite{b40}}
	\label{fig_15}
\end{figure}
\begin{table}[!t]
	\renewcommand{\arraystretch}{1.3}
	\setlength\tabcolsep{5pt}
	\caption{Performance Comparison Under Different Metrics For Molecular Dynamics Code DAG}
	\label{table2}
	\centering
	\begin{tabular}{|c|c|c|c|c|}
		\hline
		\textbf{Metric}                 & \textbf{HEFT} & \textbf{MGA} & \textbf{LA} & \textbf{RFID} \\ \hline
		Overall completion time          & 7.7240        & 7.0638       & 6.8537             & 6.2233        \\ \hline
		Execution success rate & 72.3          & 62.7         & 72.7               & 100           \\ \hline
		Algorithm running time  & 0.0499        & 0.6575       & 0.1409             & 0.1268        \\ \hline
	\end{tabular}
\end{table}

\subsection{Simulation Results for Real Application DAG Task}
Fig. \ref{fig_15} depicts a real-world DAG task of a modified molecular dynamic code \cite{b27},\cite{b40}. Table \ref{table2} presents the performance comparison of different algorithms regarding the overall DAG task completion time, the execution success rate, and the algorithm running time. It can be observed that LA and RFID algorithms exhibit a higher running time than others (i.e., HEFT and MGA). However, regarding overall DAG task completion time, the performance improvement of RFID is 19.43$\%$ better than HEFT, 11.9$\%$ better than MGA, 9.19$\%$ better than LA, and is 27.7$\%$ better than HEFT, 37.3$\%$ better than MGA, 27.3$\%$ better than LA in terms of execution success rate. In summary, simulation results verify that our proposed RFID algorithm offers an efficient and commendable reference in scheduling DAG tasks over dynamic VCs. 

\section{Conclusion}
In this paper, we investigate the DAG task scheduling problem over dynamic VC with the goal of  minimizing the overall DAG task completion time while ensuring high execution success rate. We formulate DAG task scheduling as a 0-1 integer programming problem, which is NP-hard. To tackle the problem, we propose RFID, which considers the availability and scarcity of vehicles' resources in determining the scheduling priority of different subtasks. Subsequently, RFID selects the processing vehicles based on their degree (i.e., feasible V2V connections) and resources to reduce the latency of task processing while ensuring a high reliability. Comprehensive simulations are conducted to evaluate the performance of the proposed RFID while considering multiple existing benchmarks for performance comparison. Performance comparisons reveal that our proposed RFID outperforms the existing methods in terms of the overall DAG task completion time and the execution success rate while having a marginally higher running time. Several future directions can be considered, such as the cooperation among different VCs, resource competitions among multiple DAG takes, and auction-based task allocation mechanisms.

\end{document}